\documentclass[twocolumn]{pasj01}
\usepackage{lscape} 
\Received{$\langle$2022 December 27$\rangle$}
\Accepted{$\langle$2023 March 14$\rangle$}
\Published{$\langle$2023 April 8$\rangle$}

\usepackage{setspace}

\begin{document}

\title{Multicolor and multi-spot observations of Starlink's Visorsat}
\author{Takashi H{\footnotesize ORIUCHI}$,^{1,2,*}$ Hidekazu H{\footnotesize ANAYAMA},$^2$ Masatoshi O{\footnotesize HISHI},$^{3,4}$ 
Tatsuya N{\footnotesize AKAOKA},$^{5}$ Ryo I{\footnotesize MAZAWA},$^{6}$ Koji S. K{\footnotesize AWABATA},$^{5}$ Jun T{\footnotesize AKAHASHI},$^{7}$ Hiroki O{\footnotesize NOZATO},$^{8,7}$ 
Tomoki S{\footnotesize AITO},$^{7}$ Masayuki Y{\footnotesize AMANAKA},$^{9}$ Daisaku N{\footnotesize OGAMI},$^{10}$ Yusuke T{\footnotesize AMPO},$^{10}$ Naoto K{\footnotesize OJIGUCHI},$^{10}$ 
Jumpei I{\footnotesize TO},$^{10}$ Masaaki S{\footnotesize HIBATA},$^{10}$ Malte S{\footnotesize CHRAMM},$^{11}$ Yumiko O{\footnotesize ASA},$^{12,13,14}$ Takahiro K{\footnotesize ANAI},$^{12}$ 
Kohei O{\footnotesize IDE},$^{13}$ Katsuhiro L. M{\footnotesize URATA},$^{15}$ Ryohei H{\footnotesize OSOKAWA},$^{15}$ Yutaka T{\footnotesize AKAMATSU},$^{15}$ 
Yuri I{\footnotesize MAI},$^{15}$ Naohiro I{\footnotesize TO},$^{15}$ Masafumi N{\footnotesize IWANO},$^{15}$ Seiko T{\footnotesize AKAGI},$^{16}$ Tatsuharu O{\footnotesize NO}$^{16}$, and 
Vladimir V. K{\footnotesize OUPRIANOV}$^{17}$}%

\altaffiltext{1}{Institute of Astronomy, Graduate School of Science, University of 
Tokyo, 2-21-1, Osawa, Mitaka, Tokyo 181-0015, Japan}
\altaffiltext{2}{Ishigakijima Astronomical Observatory, 
National Astronomical Observatory of Japan, 
National Institutes of Natural Sciences, 
1024-1 Arakawa, Ishigaki, Okinawa, 907-0024, Japan}
\altaffiltext{3}{Spectrum Management Office, Public Relations Center, 
National Astronomical Observatory of Japan, 
2-21-1 Osawa Mitaka Tokyo 181-8588, Japan}
\altaffiltext{4}{Department of Astronomical Science, 
SOKENDAI (The Graduate University for Advanced Studies), 
2-21-1 Osawa, Mitaka, Tokyo 181-8588, Japan}
\altaffiltext{5}{Hiroshima Astrophysical Science Center, Hiroshima University, Higashi-Hiroshima, Hiroshima 739-8526, Japan}
\altaffiltext{6}{Department of Physics, Graduate School of Advanced Science and Engineering, Hiroshima University, 
Kagamiyama, 1-3-1 Higashi-Hiroshima, Hiroshima 739-8526, Japan}
\altaffiltext{7}{Nishi-Harima Astronomical Observatory, Center for Astronomy, University of Hyogo, 407-2 Nishigaichi, Sayo-cho, Sayo, Hyogo 679-5313, Japan}
\altaffiltext{8}{Astronomy Data Center, National Astronomical Observatory of Japan, 2-21-1 Osawa Mitaka Tokyo 181-8588, Japan}
\altaffiltext{9}{Amanogawa Galaxy Astronomy Research Center (AGARC),
Graduate School of Science and Engineering, Kagoshima University, Kagoshima 890-0065, Japan}
\altaffiltext{10}{Department of Astronomy, Kyoto University, Kitashirakawa-Oiwakecho, Sakyo-ku, Kyoto 606-8502, Japan}
\altaffiltext{11}{Universität Potsdam, Karl-Liebknecht-Str. 24/25, D-14476 Potsdam, Germany}
\altaffiltext{12}{Graduate School of Science and Engineering, Saitama University, 255 Shimo-Okubo, Sakura-ku, Saitama, 338-8570}
\altaffiltext{13}{Graduate school of Education, Saitama University, 255 Shimo-Okubo, Sakura-ku, Saitama, 338-8570}
\altaffiltext{14}{Faculty of Education, Saitama University, 255 Shimo-Okubo, Sakura-ku, Saitama, 338-8570}
\altaffiltext{15}{Department of Physics, Tokyo Institute of Technology, 2-12-1 Ookayama, Meguro-ku, Tokyo 152-8551}
\altaffiltext{16}{Department of Cosmosciences, Graduate School of Science, Hokkaido University, Kita 10 Nishi8, Kita-ku, Sapporo 060-0810, Japan}
\altaffiltext{17}{Department of Physics and Astronomy, University of North Carolina at Chapel Hill, 120 E Cameron Ave, Phillips Hall CB\#3255, Chapel Hill, NC 27599-3255, USA}

\email{t-horiuchi@ioa.s.u-tokyo.ac.jp}


\maketitle

\begin{abstract}
This study provides the results of simultaneous multicolor observations 
for the first Visorsat (STARLINK-1436) and the ordinary Starlink satellite, 
STARLINK-1113 in the $U$, $B$, $V$, $g'$, $r$, $i$, $R_{\rm C}$, $I_{\rm C}$, 
$z$, $J$, $H$, and $K_s$ bands  to quantitatively investigate the extent to which 
Visorsat reduces its reflected light. Our results are as follows: (1) 
in most cases, Virorsat is fainter than STARLINK-1113, and the sunshade on Visorsat, 
therefore, contributes to the reduction of the reflected sunlight; (2) the magnitude at 550 km 
altitude (normalized magnitude) of both satellites often reaches the naked-eye limiting 
magnitude ($<$ 6.0); (3) from a blackbody radiation model of the reflected flux, the 
peak of the reflected components of both satellites is around the $z$ band; and (4) the 
albedo of the near infrared range is larger than that of the optical range. Under the assumption 
that Visorsat and STARLINK-1113 have the same reflectivity, we estimate the covering 
factor, $C_{\rm f}$, of the sunshade on Visorsat, using the blackbody radiation model: 
the covering factor ranges from $0.18 \leq C_{\rm f} \leq 0.92$. From the multivariable 
analysis of the solar phase angle (Sun-target-observer), the normalized magnitude, 
and the covering factor, the phase angle versus covering factor distribution 
presents a moderate anti-correlation between them, suggesting that the 
magnitudes of Visorsat depend not only on the phase angle but also on the orientation 
of the sunshade along our line of sight. However, the impact on astronomical 
observations from Visorsat-designed satellites remains serious. Thus, new 
countermeasures are necessary for the Starlink satellites to further reduce reflected 
sunlight.
\end{abstract}

\section{ Introduction }
Several satellite operators have developed or are developing so-called 
mega-constellation projects to enhance the satellite internet access. 
On 2019 May 24, the first 60 Starlink satellites were 
launched to a low Earth orbit (LEO) by the US satellite operator SpaceX (United 
States $[$US$]$). SpaceX planned to have launched 42,000 Starlink LEO 
communication satellites by the mid-2020s. The International Astronomical Union (IAU) 
declared its concern that the extremely bright sunlight reflected from 
the Starlink satellites would affect the pristine appearance of the dark sky 
and astronomical observations\footnote{$\langle$https://www.iau.org/news/pressreleases/detail/iau2001/$\rangle$}. 
SpaceX has attempted to reduce the reflectivity of Starlink 
satellites and has asked astronomers to evaluate satellite 
brightness in response to the concerns of the IAU. The magnitudes of Starlink 
satellites and their impact on astronomical observations have been
reported since the launch of the satellites (\cite{2020A&A...636A.121H}; 
\cite{2020arXiv200307805M}; \cite{2020ApJ...892L..36M}; \cite{2020A&A...637L...1T}, 
2021). Namely, it has been suggested that (1) in the optical and IR wavelength 
ranges on telescopes, wide-field imaging surveys would be significantly affected 
by saturation produced by the mega-constellations of LEO satellites and ghosting 
from a satellite, and (2) those satellites will have a negative impact on 
observations with a long exposure and/or a wide field of view (FoV) in the twilight. 
According to the estimation by \citet{2021MNRAS.504L..40K}, 
the additional light pollution from mega-constellations and/or other 
large sets of orbiting bodies increases the luminance of a natural night 
sky by approximately 10$\%$. 

On 2020 January 7, SpaceX launched the third batch of 60 LEO 
satellites. Among the 60 satellites, one satellite, STARLINK-1130 
(nicknamed Darksat), is a prototype with a darkening treatment on its 
communication antenna to reduce the sunlight reflected to the 
Earth. The magnitudes of Darksat and STARLINK-1113 (i.e., one 
of the normal Starlink satellites) have been evaluated. Consequently, 
Darksat is 0.77 (for the Sloan $g'$ band; \cite{2020A&A...637L...1T}) 
and has a 0.42 $-$ 0.35 magnitude dimmer (for $K_s$ and $J$ band; see 
Table 3 of \cite{2021A&A...647A..54T}) than STARLINK-1113 did. 
\citet{2020ApJ...905....3H} estimated the SDSS $g'$, the 
Cousins $R_{\rm C}$, and $I_{\rm C}$ (hereafter, $g'$, $R_{\rm C}$, and 
$I_{\rm C}$, respectively) using simultaneous observations 
from the Murikabushi telescope/MITSuME system; they concluded  
that the coating on Darksat reduces its reflectivity by approximately half that 
of STARLINK-1113. However, the apparent magnitude of Darksat 
is $\gtrsim$ 6.5 mag in the optical range, and its impact on astronomical 
observations is serious because the darkening treatment on 
Darksat alone is insufficient to abate reflected sunlight.  

In addition to Darksat, SpaceX developed Starlink Visorsat with a sun 
visor to reduce the sunlight reflection. The first Visorsat, STARLINK-1436, 
was launched on 2020 June 3. All of the Starlink satellites launched in August 
were Visorsat-designed satellites. \citet{2021arXiv210100374M} summarized 
the magnitudes of Visorsats in the visual band, and concluded that the mean 
visual magnitudes of Visorsats is 5.92 $\pm$ 0.04 mag at an orbital height 
of 550 km. According to the measurements with the Dominion Astrophysical 
Observatory’s Plaskett 1.8 m telescope via the $g'$ bandpass, the median 
magnitudes equivalent to the 550 km altitude of Visorsat and the ordinary 
Starlink satellites were 5.7 mag and 5.1 mag, respectively \citep{2022AJ....163..199B}.
However, multi-wavelength magnitudes of Visorsat have not yet 
been reported. 

In this study, we verify the effectiveness of the sun visor of STARLINK-1436 
(hereafter, Visorsat) is, thorough the UV/optical-to-NIR observations of  
Visorsat and STARLINK-1113 with the observation collaboration of optical and 
infrared synergetic telescopes for education and research (OISTER; M. Yamanaka 
in preparation), a collaboration project among Japanese universities. In section 2, 
we present the observations and data analysis of Visorsat and STARLINK-1113. 
In section 3, we describe the apparent and orbital altitude-scaled magnitudes of 
the satellites. In section 4, we discuss the albedo of Visorsat in each band and 
the shading effect of the sun visor.

\section{ Observations with the OISTER collaboration and Analysis }

\subsection{ Observations }
We conducted UV/optical-to-NIR imaging observations of Visorsat 
and STARLINK-1113. Table 1 lists the telescopes, instruments, color 
filters, and FoVs used in this study via the OISTER collaboration. Among the 
instruments in table 1, four conduct simultaneous multicolor observations: 
MITSuME ($g'$, $R_{\rm C}$, and $I_{\rm C}$ bands), HONIR (one optical 
and one NIR bands), NIC ($J$, $H$, and $K_s$ bands), and MuSaSHI ($r$, 
$i$, and $z$ bands). The equatorial coordinates of the satellites were 
forecasted by $\tt{Heavensat}$\footnote{$\langle$http://www.sat.belastro.net/heavensat.ru/english/index.html$\rangle$} 
using the two-line element (TLE) data from the Celestrak 
website\footnote{$\langle$https://celestrak.com/satcat/search.php$\rangle$}.
Tracking the Starlink satellites with the telescopes listed in table 1 is difficult  
because these satellites move at high speeds (e.g., over 1000 arcsec s${^{-1}}$). 
Instead, we observed satellite trails and bright reference stars using star tracking. 
Figures 1 and 2 show examples of the trail images of Visorsat and STARLINK-1113 
captured with the OISTER collaboration. The star images were elongated by stopping the 
telescopes to unify the photometric method of the satellite trails and comparison stars. 
However, not all of the telescopes in this study were able to elongate object images. 
Therefore, we measured the satellite magnitudes of the two methods as shown in 
the next section. 

Table 2 summarizes the observation logs used in this study for the measurement of 
the satellite magnitudes and various analyses. We used orbital information (e.g., 
the coordinates at exposure times, phase angles, and distance between the satellite 
and observer) from the HORIZONS Web-Interface\footnote{$\langle$https://ssd.jpl.nasa.gov/horizons.cgi$\rangle$}. 
The exposure times were set to 5 s (or 3 s) to increase the success rate of 
observations and to improve the signal-to-noise ratio of reference stars, 
referring to \citet{2020ApJ...905....3H}. The exception is the UV/optical bands of 
the Kanata telescope/HONIR, which has a 20 s exposure owing to the constraint 
of the observation system (see also subsection 3.2).

\begin{figure*}
   \includegraphics[height=16cm,width=16cm]{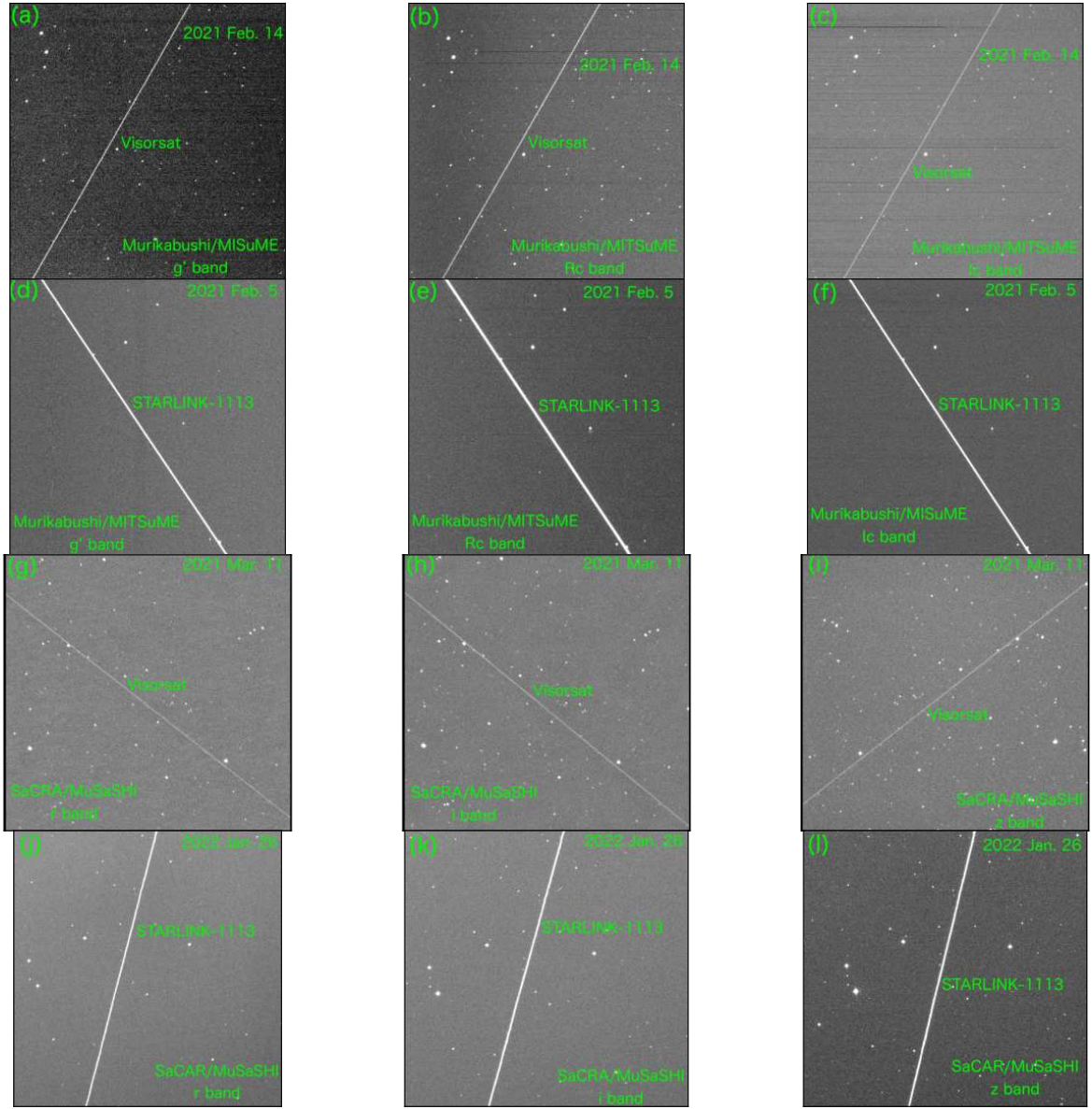} 
\caption{Examples of images of Visorsat (a, b, and c: 
Murikabushi/MITSuME $g'$, $R_{\rm C}$, and $I_{\rm C}$ bands, g, h, and i: 
SaCRA/MuSaSHI $r$, $i$ and $z$ bands) and STARLINK-1113 trails (d, e, and 
f: Murikabushi/MITSuME $g'$, $R_{\rm C}$, and $I_{\rm C}$ bands, j, k, and l: 
SaCRA/MuSaSHI r, i, and z bands). Details of the observations (date, target, 
and telescope/instrument) are described in each panel.}
\end{figure*}

\begin{figure*}
   \includegraphics[height=14cm,width=16cm]{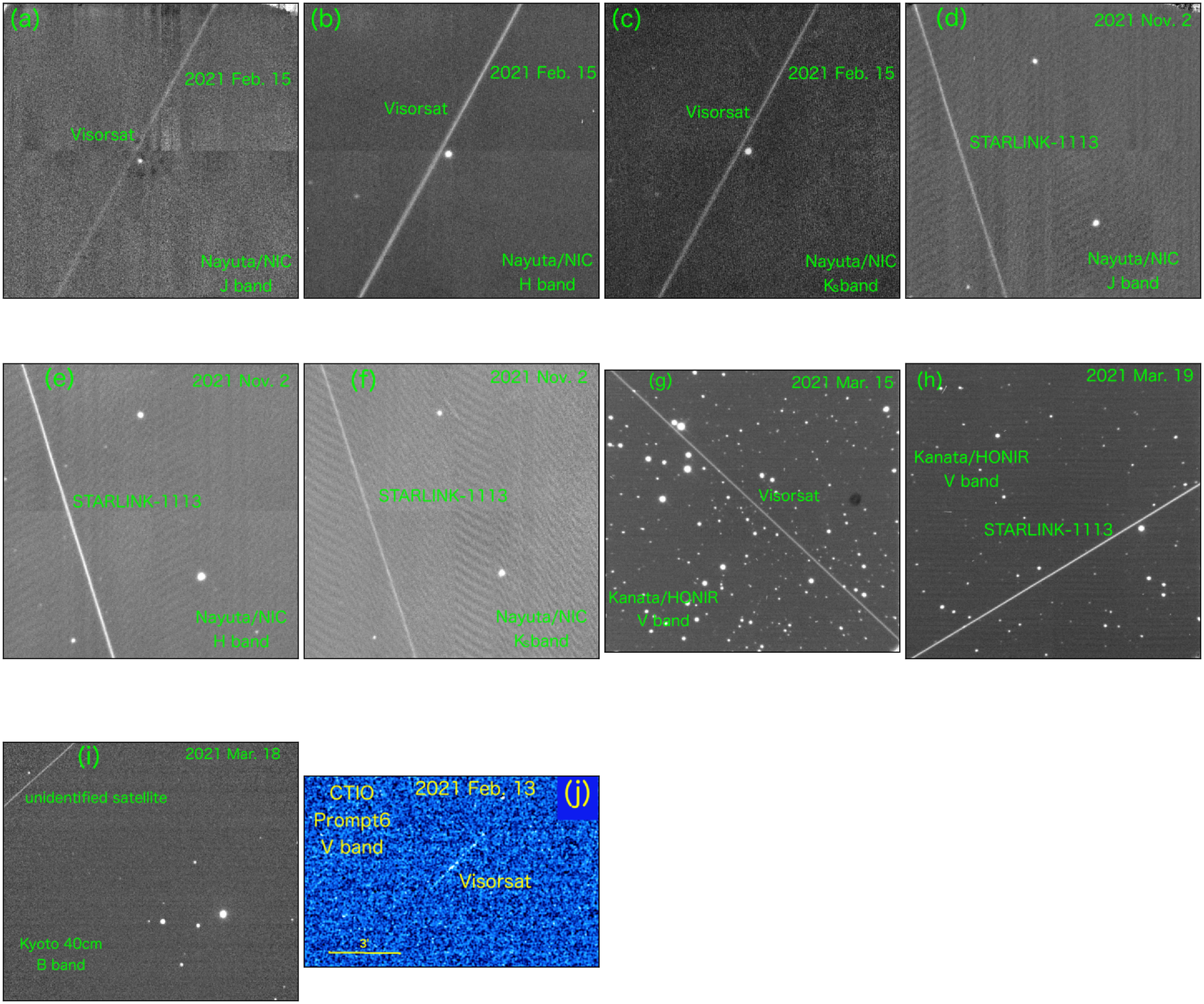} 
\caption{Same as figure 1, but for the Nayuta/NIC, Kanata/HONIR, 40 cm 
telescope/Visual CCD, and PROMPT-6 (Skynet/CTIO) images.
The image captured with the 40cm telescope (or PROMPT-6) is inverted 
(or rotated) from north to south (or by 90$^\circ$ clockwise).  The 
PROMPT-6 data (panel j) are from a test image with a 0.2 s exposure.} 
\end{figure*}

\subsection{ Data Analysis }
We conducted dome flat corrections and subtraction of dark, sky, and bad pixels 
and performed aperture photometry on the data using the Image Reduction and 
Analysis Facility package. Star images or bad pixels in the satellite trails were 
subtracted using images of the same target fields without these trails. We 
did not use the data taken with the 40 cm telescope and 0.41 m PROMPT-6 from our 
analysis, owing to their data having low signal-to-noise ratios $S/N\sim1$).

\subsubsection{ With Elongated Star Images }
In this sub-subsection, we explain the method for measuring the satellite flux and 
magnitudes, referring to that described in \citet{2020ApJ...905....3H}.
Because the satellite flux, $f_{\rm sat}$, is diluted by the angular velocity of the 
satellites, $V_{\rm sat}$ (i.e., $f_{\rm sat}\propto V_{\rm sat}^{-1}$), the magnitude 
of the satellite must consider the angular velocity as follows: 
\begin{eqnarray}
m_{\rm sat} &=& m_{\rm star} - 2.5\log\biggr(\frac{f_{\rm sat}}{f_{\rm star}}\frac{V_{\rm sat}}{V_{\rm star}}\biggr), 
\end{eqnarray}
where $V_{\rm star}$ ($=~15~\cos\delta$ arcsec s$^{-1}$, $\delta:$ decl.), 
$m_{\rm star}$, and $f_{\rm star}$ are the angular velocities along celestial 
sphere, magnitude, and observed flux of the reference stars, respectively. 
In this case, $f_{\rm star}$ is the reference star flux of the elongated images.
In this manner, when both satellites and stars are elongated images, the scaling 
by the velocity ratio, $V_{\rm sat}/V_{\rm star}$, is necessary because either of 
those images is not the beginning and/or the end within the FoV. The angular 
velocity of the satellites $V_{\rm sat}$ is expressed as the transverse speed 
at the great-circular distance, $\lambda$:
\begin{eqnarray}
\lambda = \arccos{(\sin\delta\sin D + \cos\delta\cos D\cos(A-\alpha))},
\end{eqnarray}
where $(\alpha,~\delta)$ and $(A,~D)$ are the right ascension, declination at the 
initial time and that at a certain time of observations, respectively. The magnitude 
uncertainties, $\sigma_{\rm m}$, of the satellite trails can be estimated 
by the law of error propagation:   
\begin{eqnarray}
\sigma_m &=& \frac{2.5}{\ln 10}\sqrt{\biggr(\frac{\delta f_{\rm sat}}{f_{\rm sat}}\biggr)^2+\biggr(\frac{\delta f_{\rm star}}{f_{\rm star}}\biggr)^2+\biggr(\frac{\delta V_{\rm sat}}{V_{\rm sat}}\biggr)^2}, 
\end{eqnarray}
where $\delta f_{\rm sat}$, $\delta f_{\rm star}$, and $\delta V_{\rm sat}$ are the 
flux errors for the satellite trails, comparison stars, and the angular-velocity 
error of satellites, respectively. The flux error of the satellite trails $\delta f_{\rm sat}$ 
or that of comparison stars $\delta f_{\rm star}$ can be estimated by the standard 
deviation of sky flux, $\sigma_{\rm sky}$, around the streaks:  
$\delta f_{\rm sat}$ (or $\delta f_{\rm star}$) is the $\sigma_{\rm sky}~\times~$ 
pixel width of the satellite (or elongated-star) trails. The angular-velocity error of a 
satellite is the mean value of the velocity difference between the central to the start 
exposure time, $|V_{\rm sat}~-~V_{\rm start}|$, and that of the central to the end time, 
$|V_{\rm end}~-~V_{\rm sat}|$. 

Figure 3 presents the pixel-wide average cross-sectional profiles of the Visorsat and 
STARLINK-1113 streaks captured by the Murikabushi telescope/MITSuME. 
These profiles were measured by $\tt{Projection}$ of SAOimage DS9\footnote{$\langle$https://sites.google.com/cfa.harvard.edu/saoimageds9$\rangle$}, 
which allowed us to evaluate the average count of the cross-section sums of 
satellite trails along a rectangular region. We applied the following 
procedure to display the cross section of the satellite flux using $\tt{Projection}$: (1) 
make a vertical incision into the satellite trail with the width of the trail; (2) 
extend the incised area parallel to the thickness of the satellite trail, and 
create a rectangular area. Therefore, the average cross-sectional profile 
of the trail appears. We evaluated the satellite flux, $f_{\rm sat}$, using 
$\sum_{i=-m}^m\bar{F_i}$, where $\bar{F_i}$ is the count of the cross-section
profile at the $i$-th pixel, and its pixel width is $2m+1$. We applied the same 
rectangular shape as that used for the estimation of $f_{\rm sat}$ or $f_{\rm star}$ 
to the sky region used to evaluate $\sigma_{\rm sky}$.

We elongated star images around target fields by tracking-off observations, so 
as to compare the flux counts of the Starlink satellites with those of reference stars in 
the same manner as that of the satellite trails (i.e., to apply equation 1). 
The elongated star images were obtained with the Murikabushi telescope, 
Akeno 50 cm telescope/MITSuME, and the Kanata telescope/HONIR. 
We selected the first and/or second brightest stars in the FoV as reference stars, 
which did not merged with nearby stars. Because both edges of the elongated-star 
images are rounded, they are not used when measuring the flux by 
$\tt{Projection}$. For details on the methods using elongated stars, refer to 
\citet{2020ApJ...905....3H}.

\subsubsection{ Without Elongated Star Images }
For the cases in which elongated star images could not be obtained, the photometry of the 
satellites was performed by comparing the flux of satellite streaks with that of point 
sources (i.e., comparison stars). The total flux of a satellite trail is represented by $f_{\rm sat}L$, 
where $L$ is the length of the trail in the FoV. The definition of $f_{\rm sat}$ is the 
same as that described in the previous sub-subsection. Using the flux of comparison 
stars, $f_{\rm exp}$, at the exposure time, $t_{\rm exp}$, the flux at the effective satellite 
exposure time, $t_{\rm eff}~(=~L/V_{\rm sat})$, is described as $f_{\rm exp}(t_{\rm eff}/t_{\rm exp})$. In other words, the satellite magnitude, 
\begin{eqnarray}
\nonumber m_{\rm sat} &=& m_{\rm star} - 2.5\log\biggr[\frac{f_{\rm sat}L}{f_{\rm exp}(t_{\rm eff}/t_{\rm exp})}\biggr] \\
			            &=&  m_{\rm star} - 2.5\log\biggr(\frac{f_{\rm sat}V_{\rm sat}t_{\rm exp}}{f_{\rm exp}}\biggr),
\end{eqnarray}
does not depend on the trail length, $L$. In this case, the magnitude error of the 
satellites is expressed as follows:
\begin{eqnarray}
\sigma_m &=& \frac{2.5}{\ln 10}\sqrt{\biggr(\frac{\delta f_{\rm sat}}{f_{\rm sat}}\biggr)^2+\biggr(\frac{\delta f_{\rm exp}}{f_{\rm exp}}\biggr)^2+\biggr(\frac{\delta V_{\rm sat}}{V_{\rm sat}}\biggr)^2}, 
\end{eqnarray}
where $\delta f_{\rm exp}$ is the flux error of the comparison stars measured using 
aperture photometry. Notably, the photometric method 
using equation (1) would not be more advantageous or disadvantageous than the 
methods described in this sub-subsection.

\subsubsection{ Magnitude conversion }
To evaluate the multicolor magnitude of Visorsat and STARLINK-1113, 
we adopt the UCAC4 catalog, which is able to refer to the SDSS $r$, $i$, 
Johnson's $B$, $V$, $J$, and $K_s$ band magnitudes. The $B$, $V$, $r$, 
and $i$ band magnitudes can be converted to the $g'$, $R_{\rm C}$, 
$I_{\rm C}$ (for MITSuME), and $z$ band magnitudes (for MuSaSHI) by using 
the color terms and zeropoints in table 3 of \citet{2006A&A...460..339J}:
\begin{eqnarray}
g' &=& V + 0.630(B - V) - 0.124,\\
R &=& r - 0.252(r - i) - 0.152, \\
I &=& r - 1.245(r - i) - 0.387, \\
z &=& r - 1.573(r - i) + 0.012.
\end{eqnarray}
Because the $H$ band magnitudes are not listed in the UCAC4 catalog, we 
referred to them in the 2MASS data. We confirmed that the $J$ and $K_s$ 
band magnitudes in the UCAC4 catalog were the same values as those in 
2MASS. 

\begin{table*}
\tbl{The list of telescopes/Instrument, filters, and references for this observation collaboration.}{%
\begin{tabular}{cccccccccc}
\hline\noalign{\vskip3pt}
Institution & Telescope & Instrument & Filter(s)$^*$ & FoV & Ref.   \\   
\hline  
IAO/NAOJ & 105 cm Murikabushi telescope & MITSuME & $g',~R_{\rm C},~I_{\rm C}$  & \timeform{12'.3} $\times$ \timeform{12'.3} & (1), (2), (3) \\
Hiroshima University & 1.5 m Kanata telescope & HONIR & $B,~V,~H$ & \timeform{10'} $\times$ \timeform{10'} & (4) \\
University of Hyogo & 2.0 m Nayuta telescope & NIC & $J,~H,~K_s$ & \timeform{2'.7} $\times$ \timeform{2'.7} & (5) \\
Kyoto University & 40 cm Schmidt-Cassegrain telescope & Visual CCD & $B$ & \timeform{11'.5} $\times$ \timeform{11'.5} & --- \\
Akeno Observatory & 50 cm telescope Akeno & MITSuME & $g',~R_{\rm C},~I_{\rm C}$ & \timeform{28'} $\times$ \timeform{28'} & (1), (2), (3) \\
Saitama University & 0.55 m SaCRA telescope & MuSaSHI & $r,~i,~z$ & \timeform{12'.8} $\times$ \timeform{12'.4} & (6)\\
Hokkoido University & 1.6 m Pirka telescope & MSI & $U$ &  \timeform{3'.3} $\times$ \timeform{3'.3} & (7) \\
Skynet/CTIO & 0.41 m PROMPT-6 & FLI PL23042 & $V$ & \timeform{15'.1} $\times$ \timeform{15'.1} & (8) \\
\hline\noalign{\vskip3pt} 
\end{tabular}}\label{table:extramath}
\begin{tabnote}
\hangindent6pt\noindent
\hbox to6pt{\footnotemark[$*$]\hss}\unskip
Filters used in this study.\\
References $-$ (1) \citet{2005NCimC..28..755K},  (2) \citet{2007PhyE...40..434Y}, (3) \citet{2008AIPC.1000..543S}, 
(4) \citet{2014SPIE.9147E..4OA}, (5) \citet{2014PASJ...66...53T}, (6) \citet{2020SPIE11447E..5ZO}, 
(7) \citet{2012SPIE.8446E..2OW}, and (8) \citet{2022PASP..134a5001D}
\end{tabnote}
\end{table*}

\begin{spacing}{0.8}
\begin{table*}
\tbl{Observation log of Visorsat and STARLINK-1113 for usable data. \label{tab:mathmode}}{
\begin{tabular}{ccccc}
\hline\noalign{\vskip3pt}
Target & Visorsat & Visorsat & Visorsat & STARLINK-1113 \\
Observation Date & 2021 Feb 14 & 2021 Feb 15 & 2021 Mar 14 & 2021 Feb 5 \\ 
Telescope&Murikabushi &Murikabushi&Murikabushi&Murikabushi\\
Instrument &MITSuME  &MITSuME&MITSuME&MITSuME\\
Filter(s) & $g',~R_{\rm C},~I_{\rm C}$ & $g',~R_{\rm C},~I_{\rm C}$ & $g',~R_{\rm C},~I_{\rm C}$ & $g',~R_{\rm C},~I_{\rm C}$\\
Start Time of Observation (UTC) & 21:40:57 & 21:33:27 & 12:03:17 & 11:10:27   \\ 
Central Time of Observation (UTC) & 21:41:00 & 21:33:30 & 12:03:20 & 11:10:30 \\ 
End Time of Observation (UTC) & 21:41:02 & 21:33:32  & 12:03:22 & 11:10:32\\ 
Exposure Time (s) & 5.0 & 5.0 & 5.0 & 5.0 \\
RA &  \timeform{17h53m52.5s} & \timeform{14h13m33.94s} & \timeform{02h35m05.7s}  &  \timeform{02h45m14.2s}  \\ 
Dec & +\timeform{02D04'06.8''} & +\timeform{21D39'01.7''} & +\timeform{34D23'55.4''} & $-$\timeform{16D09'41.9"} \\  
Azimuth ($^\circ$) & 61.6 & 265.3 & 299.0 & 214.7 \\ 
Elevation ($^\circ$) & 50.3 & 71.4 & 22.8 & 42.2 \\ 
Airmass & 1.30 & 1.05 & 2.55 & 1.48 \\ 
$D^*$ (km) & 694.54 & 575.51 & 1189.77 & 781.42  \\ 
Solar Phase Angle ($^\circ$) & 118.7 & 61.4 & 124.4 & 101.5 \\
Observer Phase Angle ($^\circ$) & 61.3 & 118.6 & 55.6 & 78.4 \\  
Angular Velocity (arcsec s$^{-1}$) & 1826.78 $\pm$ 49.04 & 2587.01 $\pm$ 26.01 & 812.41 $\pm$ 16.89 & 1417.02 $\pm$ 41.15 \\
\hline
Target & STARLINK-1113 & STARLINK-1113 & STARLINK-1113 & Visorsat \\
Observation Date & 2021 Mar 17 & 2021 Mar 18 & 2021 Mar 22 & 2021 Nov 11  \\ 
Telescope& Akeno 50 cm & Akeno 50 cm & Akeno 50 cm & Akeno 50 cm\\
Instrument & MITSuME & MITSuME & MITSuME & MITSuME \\
Filter(s) & {\bf $g',~R_{\rm C},~I_{\rm C}$} & {\bf $g',~R_{\rm C},~I_{\rm C}$} & {\bf $g',~R_{\rm C},~I_{\rm C}$} & {\bf $g',~R_{\rm C},~I_{\rm C}$} \\
Start Time of Observation (UTC) & 19:53:58 & 19:48:28 & 19:26:58 & 09:12:28 \\ 
Central Time of Observation (UTC) & 19:54:00 & 19:48:30 & 19:27:00 & 09:12:30 \\ 
End Time of Observation (UTC) & 19:54:03 & 19:48:33 & 19:27:03 & 09:12:33 \\ 
Exposure Time (s) & 5.0 & 5.0 & 5.0 & 5.0 \\
RA & \timeform{18h26m36.3s} & \timeform{17h27m46.3s} & \timeform{10h41m52.5s} & \timeform{23h37m43.5s} \\ 
Dec & \timeform{-02D11'27.6''} &  +\timeform{14D02'19.3''} & +\timeform{66D20'40.9''} & \timeform{-18D41'37.5''} \\  
Azimuth ($^\circ$) & 143.8 & 155.1 & 331.5 & 149.9 \\ 
Elevation ($^\circ$) & 45.8 & 66.5 & 32.4 & 29.8 \\ 
Airmass & 1.40 & 1.09 & 1.86 & 2.08 \\
$D^*$ (km) & 781.42 & 595.91 & 944.37 & 993.21 \\
Solar Phase Angle ($^\circ$) & 99.3 & 83.7 & 68.9 & 62.6 \\
Observer Phase Angle ($^\circ$) & 80.7 & 96.3 & 111.1 & 117.4 \\  
Angular Velocity (arcsec s$^{-1}$) & 2021.02 $\pm$ 13.43 & 2510.70 $\pm$ 19.97 & 1582.80 $\pm$ 5.74& 1461.82 $\pm$ 14.7 \\
\hline\noalign{\vskip3pt} 
\end{tabular}}\label{table:extramath}
\begin{tabnote}
\hangindent6pt\noindent
\hbox to6pt{\footnotemark[$*$]\hss}\unskip
Distance between Satellite and Observer.\\
\end{tabnote}
\end{table*}
\end{spacing}

\begin{table*}
\setcounter{table}{1}
\tbl{(Continued) \label{tab:mathmode}}{
\begin{tabular}{ccccc}
\hline\noalign{\vskip3pt}
Target & Visorsat & Visorsat & Visorsat & Visorsat  \\
Observation Date & 2022 Jan 17 & 2021 Mar 15 & 2021 Mar 15 & 2022 Jan 17 \\ 
Telescope&Akeno 50 cm & Kanata & Kanata & Kanata\\
Instrument & MITSuME & HONIR  & HONIR  & HONIR \\
Filter(s) & {\bf $g',~R_{\rm C},~I_{\rm C}$} & {\bf $V$} & {\bf $H$} & {\bf $V$} \\
Start Time of Observation (UTC) & 9:27:58 & 10:19:52 & 10:19:52 & 9:26:54 \\ 
Central Time of Observation (UTC) & 9:28:00 & 10:20:02 & 10:19:55 & 9:27:04 \\ 
End Time of Observation (UTC) & 9:28:03 & 10:20:12 & 10:19:57 & 9:27:14 \\ 
Exposure Time (s) & 5.0 & 20.0 & 5.0 & 20.0 \\
RA & \timeform{00h06m52.27s} & \timeform{07h00m34.15s} & \timeform{06h49m6.8s} &  \timeform{01h12m24.5s} \\ 
Dec & \timeform{-10D22'29.6"} & \timeform{-04D07'00.6"} & \timeform{-01D26'45.7"} & \timeform{-19D19'29.9"} \\  
Azimuth ($^\circ$) & 222.4 & 173.0 & 177.4 & 195.0 \\ 
Elevation ($^\circ$) & 34.0 & 51.3 & 54.1 & 35.0 \\ 
Airmass & 1.63 & 1.26  & 1.26 & 1.79 \\ 
$D^*$ (km)  & 906.24 & 685.40 &  662.76 & 937.95 \\ 
Solar Phase Angle ($^\circ$) & 118.2 & 70.3 & 73.0 & 106.6 \\
Observer Phase Angle ($^\circ$) & 61.8 & 109.7 & 107.0 & 73.3 \\  
Angular Velocity (arcsec s$^{-1}$) & 1080.83 $\pm$ 29.46 & 1963.23 $\pm$ 182.15 & 2080.71 $\pm$ 44.71 & 1259.71 $\pm$ 128.12\\
\hline
Target & Visorsat & Visorsat & Visorsat & STARLINK-1113 \\
Observation Date & 2022 Jan 17 & 2022 Jan 19 & 2022 Jan 19 & 2021 Mar 19 \\ 
Telescope & Kanata & Kanata & Kanata & Kanata \\
Instrument & HONIR  & HONIR  & HONIR & HONIR  \\
Filter(s) & {\bf $H$} & {\bf $B$} & {\bf $H$} & {\bf $V$} \\
Start Time of Observation (UTC) & 09:26:54 & 09:14:53 & 09:14:53 &  19:42:21  \\ 
Central Time of Observation (UTC) & 09:26:57 & 09:15:03  & 09:14:56 & 19:42:31 \\
End Time of Observation (UTC) & 09:26:59 & 09:15:13 & 09:14:58 & 19:42:41 \\ 
Exposure Time (s) & 5.0 & 20.0 & 5.0 & 20.0 \\
RA &  \timeform{01h04m2.8s} & \timeform{23h20m55.1s} &  \timeform{23h14m51.3s}  &  \timeform{18h50m00.1s}  \\ 
Dec & \timeform{-20D55'37''} & +\timeform{17D10'26.7''} & +\timeform{13D42'20''} & +\timeform{29D50'25.0"} \\  
Azimuth ($^\circ$) & 196.8 & 254.5 & 251.5 & 88.0 \\ 
Elevation ($^\circ$) & 33.0 & 50.7 & 47.5 & 58.7 \\ 
Airmass & 1.79 & 1.31 & 1.32 & 1.16 \\ 
$D^*$ (km) & 925.08 & 690.72 & 720.08 & 635.29  \\ 
Solar Phase Angle ($^\circ$) & 108.8 & 118.9 & 122.1 & 101.3 \\
Observer Phase Angle ($^\circ$) & 71.2 & 61.1 & 57.9 & 78.7 \\  
Angular Velocity (arcsec s$^{-1}$) & 1167.52 $\pm$ 29.5 & 1872.82 $\pm$ 189.03 & 1731.73 $\pm$ 46.05 & 2244.87 $\pm$ 167.29 \\
\hline\noalign{\vskip3pt} 
\end{tabular}}\label{table:extramath}
\begin{tabnote}
\hangindent6pt\noindent
\hbox to6pt{\footnotemark[$*$]\hss}\unskip
Note 1 $-$ RA, Dec, azimuth, elevation, airmass, $D$, solar phase angle, 
and observer phase angle are values at the central exposure time. \\
Note 2 $-$ the difference of the imaging system between 
the HONIR $V$ or $B$ and $H$ passbands makes the differences of 
exposure times, RA and decl. at a central exposure time, and so on.
\end{tabnote}
\end{table*}

\begin{table*}
\setcounter{table}{1}
\tbl{(Continued) \label{tab:mathmode}}{
\begin{tabular}{ccccc}
\hline\noalign{\vskip3pt}
Target & STARLINK-1113 & STARLINK-1113 & STARLINK-1113 & STARLINK-1113 \\
Observation Date & 2021 Mar 19 & 2021 Mar 22 & 2022 Jan 24  &  2022 Jan 24 \\ 
Telescope& Kanata & Kanata & Kanata & Kanata \\
Instrument & HONIR & HONIR  & HONIR & HONIR \\
Filter(s) & {\bf $H$} & {\bf $V$} & {\bf $B$} & {\bf $H$} \\
Start Time of Observation (UTC) & 19:42:21 & 19:26:42 & 10:34:55 & 10:34:55 \\ 
Central Time of Observation (UTC) & 19:42:24 & 19:26:52 & 10:35:05 & 10:34:58 \\ 
End Time of Observation (UTC) & 19:42:26 & 19:27:02 & 10:35:15 & 10:35:00 \\ 
Exposure Time (s) & 5.0 & 20.0 & 20.0 & 5.0 \\
RA & \timeform{18h31m29.2s} & \timeform{00h17m23.4s} & \timeform{23h24m30.6s} & \timeform{23h09m23.1s} \\ 
Dec & +\timeform{27D18'09.7''} &  +\timeform{83D41'59.3''} & +\timeform{42D57'26.5''} & +\timeform{43D22'14''} \\  
Azimuth ($^\circ$) & 95.6 & 6.28 & 300.2 & 301.6 \\ 
Elevation ($^\circ$) & 61.6 & 31.2 & 40.6  & 38.0 \\ 
Airmass & 1.16 & 1.90 & 1.59 & 1.59 \\
$D^*$ (km) & 618.8 & 970.8 & 805.8 & 843.3 \\
Solar Phase Angle ($^\circ$) & 97.5 & 97.1 & 105.5 & 107.0 \\
Observer Phase Angle ($^\circ$) & 82.5 & 82.9 & 74.5 & 73.0 \\  
Angular Velocity (arcsec s$^{-1}$) & 2359.22 $\pm$ 40.87 & 1281.29 $\pm$ 100.82 & 1391.29 $\pm$ 155.07 & 1279.96 $\pm$ 35.33 \\
\hline
Target & Visorsat & STARLINK-1113 & Visorsat & Visorsat  \\
Observation Date & 2021 Feb 15 & 2021 Nov 2  & 2021 Mar 11 & 2021 Mar 15 \\ 
Telescope& Nayuta & Nayuta & SaCRA & SaCRA  \\
Instrument & NIC & NIC & MuSaSHI & MuSaSHI \\
Filter(s) & {\bf $J,~H,~K_s$} & {\bf $J,~H,~K_s$} & {\bf $r,~i,~z$} & {\bf $r,~i,~z$} \\
Start Time of Observation (UTC) & 19:54:57 & 09:07:02 & 10:43:28 & 10:19:59 \\ 
Central Time of Observation (UTC) & 19:55:00 & 09:07:05 & 10:43:31 & 10:20:00 \\ 
End Time of Observation (UTC) & 19:55:02 & 09:07:07 & 10:43:33 & 10:20:02 \\ 
Exposure Time (s) & 5.0 & 5.0 & 5.0 & 3.0 \\
RA & \timeform{17h03m32.3s} & \timeform{17h13m31.0s} & \timeform{01h31m13.5s}  & \timeform{04h22m21.9s} \\ 
Dec & +\timeform{03D58'59.4"} & +\timeform{02D21'12.2"}  & +\timeform{62D03'51.9''} & \timeform{-10D10'59.8''} \\  
Azimuth ($^\circ$) & 123.7 & 249.6 & 326.4 & 229.1 \\ 
Elevation ($^\circ$) &  43.78 & 30.1 & 32.7 & 29.3 \\ 
Airmass & 1.44 & 1.99  & 1.85 & 2.03 \\
$D^*$ (km)  & 761.3 & 992.20 & 940.74 & 1005.99 \\ 
Solar Phase Angle ($^\circ$) & 105.3 & 136.0 & 110.4 & 109.7 \\
Observer Phase Angle ($^\circ$) & 74.7 & 44.0 & 69.6 & 70.3  \\  
Angular Velocity (arcsec s$^{-1}$) & 1509.96 $\pm$ 43.58 & 1027.67 $\pm$ 24.91 & 1065.11 $\pm$ 27.85  & 1506.41 $\pm$ 0.98 \\
\hline
Target & Visorsat & Visorsat & STARLINK-1113 & Visorsat  \\
Observation Date & 2021 Nov 15 & 2021 Nov 17 & 2022 Jan 26 & 2021 Mar 9 \\ 
Telescope & SaCRA & SaCRA & SaCRA & Pirka \\ 
Instrument & MuSaSHI & MuSaSHI & MuSaSHI & MSI  \\
Filter(s) & {\bf $r,~i,~z$} & {\bf $r,~i,~z$} & {\bf $r,~i,~z$} & {\bf $U$} \\
Start Time of Observation (UTC) & 08:46:59 & 08:35:00 & 08:47:59 &  09:16:57  \\ 
Central Time of Observation (UTC) & 08:47:00 & 08:35:01 & 08:48:00 & 09:17:00 \\ 
End Time of Observation (UTC) & 08:47:02 & 08:35:03 & 08:48:02 & 09:17:02 \\ 
Exposure Time (s) & 3.0 & 3.0  & 3.0 & 5.0 \\
RA & \timeform{18h21m04.2s}  & \timeform{16h22m13.1s} & \timeform{06h02m53.1s} &  \timeform{14h05m12.9s} \\ 
Dec & +\timeform{18D37''06.0'} & +\timeform{45D36'51.6''} & +\timeform{15D13'28.8''} & +\timeform{75D09'53.8''} \\  
Azimuth ($^\circ$) & 262.7 & 306.1 & 98.8 & 15.5 \\ 
Elevation ($^\circ$) & 42.5 & 32.0 & 38.0 & 35.2 \\ 
Airmass & 1.48 & 1.90 & 1.62 & 1.73 \\ 
$D^*$ (km) & 777.46 & 957.08 & 844.00 & 889.91 \\ 
Solar Phase Angle ($^\circ$) & 122.5 & 114.3 & 36.2 & 74.6 \\
Observer Phase Angle ($^\circ$) & 57.5 & 65.7 & 143.8 & 105.4 \\  
Angular Velocity (arcsec s$^{-1}$) & 1603.68 $\pm$ 23.59 & 1524.99 $\pm$ 7.85 & 1417.83 $\pm$ 20.80 & 1683.70 $\pm$ 3.95 \\
\hline
\end{tabular}}
\label{table:extramath}
\begin{tabnote}
\hangindent6pt\noindent
\end{tabnote}
\end{table*}

\begin{table*}
\setcounter{table}{1}
\tbl{(Continued) \label{tab:mathmode}}{
\begin{tabular}{ccccc}
\hline\noalign{\vskip3pt}
Target & Visorsat & STARLINK-1113 &  Visorsat & \\
Observation Date & 2021 Mar 9 & 2021 Mar 19 & 2021 Feb 13 &  \\ 
Telescope & Pirka & Pirka  & PROMPT-6 & \\ 
Instrument & MSI  & MSI  & FLI PL23042 & \\
Filter(s) & {\bf $U$} & {\bf $U$} & {\bf $V$ }&  \\
Start Time of Observation (UTC) & 10:55:07 & 19:45:17 & 00:20:00 & \\ 
Central Time of Observation (UTC) & 10:55:10 & 19:45:20  & --- & \\ 
End Time of Observation (UTC) & 10:55:12 & 19:45:22  & --- & \\ 
Exposure Time (s) & 5.0 & 5.0 & 0.2 &  \\
RA &  \timeform{02h34m12.2s} & \timeform{18h41m12.2s} & \timeform{02h50m33.6s} &  \\ 
Dec & +\timeform{32D39'18.9"} & +\timeform{19D16'29.6"} &  +\timeform{41D34'36.3"} &   \\  
Azimuth ($^\circ$) & 286.0 & 142.3 & 334.4 &  \\ 
Elevation ($^\circ$) & 32.4 & 49.1 & 11.4 &  \\ 
Airmass & 1.86 & 1.32 & 4.89 &  \\ 
$D^*$ (km) & 942.26 & 709.73 & 1719.79 &  \\ 
Solar Phase Angle ($^\circ$) & 120.7 & 100.6 & 91.16 &  \\
Observer Phase Angle ($^\circ$) & 59.3 & 79.4 & 88.83 &   \\  
Angular Velocity (arcsec s$^{-1}$) & 1120.81 $\pm$ 28.12 & 2132.09 $\pm$ 3.19 & 784.88 &  \\
\hline
\end{tabular}}
\label{table:extramath}
\begin{tabnote}
\hangindent6pt\noindent
\end{tabnote}
\end{table*}

\section{ Results }
In this section, we will present the multicolor magnitudes and colors of 
Visorsat and STARLINK-1113. 

\subsection{ Apparent and normalized magnitude }
We measured not only the apparent magnitudes but also the magnitudes 
at the satellite orbital altitude of 550 km (hereafter, normalized magnitude; see 
\cite{2020ApJ...905....3H}, \cite{2020A&A...637L...1T}, and \cite{2022AJ....163..199B}), 
because the apparent magnitude dims as the distance, $r$, between an 
observer and a satellite increases. The normalized magnitude
was calculated by adding a factor of $+5\log(550/r)$ to equations (1) and (4). We 
determined the distance error, $\delta r$: an averaged value of the distance 
difference between the start and central exposure time, $|r~-~r_{\rm start}|$, 
and that of the central to end time, $|r~-~r_{\rm end}|$. 

Table 3 presents the apparent and normalized magnitudes of Visorsat and 
STARLINK-1113, and whether equation (1) or (4) was used to 
measure the magnitudes. We confirmed the following trends: (1) the 
magnitude of Visorsat is generally dimmer than that of STARLINK-1113; (2) the 
normalized magnitudes of both satellites are often brighter ($<$ 6.0 mag) than the 
naked-eye limiting magnitude; (3) the shorter the observed wavelength is, 
the larger the satellite magnitudes tend to be. The $U$ band 
observations with the 1.6 m Pirka telescope/MSI did not detect the satellite 
trails of Visorsat nor STARLINK-1113. 

\subsection{ Phase angle dependence of the satellite magnitudes and colors }
\citet{2021arXiv210100374M} and other study\footnote{$\langle$https://amostech.com/TechnicalPapers/2021/Non-Resolved-Object-Characterization/Johnson.pdf$\rangle$} 
showed a negative (or positive) correlation between the satellite magnitude 
and phase angle for many normal Starlink satellites, Visorsats and Darksat in the 
solar phase angles of $<~90^{\circ}$ (or $>~90^{\circ}$) in the visual magnitude, 
$M_{\rm V}$. This section describes the phase angle dependence of the normalized 
magnitudes estimated in the previous sections. Figure 4 shows the phase angle 
dependence of the normalized magnitudes and colors of Visorsat and STARLINK-1113. 
The normalized magnitudes, especially for the MITSuME data (also the MuSaSHI 
data for Visorsat), tended to increase with phase angles over $\sim$ 100$^\circ$. 
We confirmed that this trend is consistent with the results of  aforementioned report even 
with multicolor observations. Because the number of data points in $<~90^\circ$ is too 
small, we cannot strictly verify the trend in this range. However, the phase angle versus 
magnitude diagrams (figures 4a and 4b) presented considerable scatters except for 
the phase angle dependence of the magnitude. For instance in figure 4a, the $g'$-, 
$R_{\rm C}$-, and $I_{\rm C}$-band magnitude differences between the datapoints 
at 61.4$^\circ$ and 62.6$^\circ$ are $\gtrsim$ 2 mag despite their small phase 
angle difference with each other. The possible causes of such magnitude scattering are 
discussed in section 4. Overall, STARLINK-1113 tended to be $\sim$ 1 mag brighter 
than Visorsat. 

To demonstrate trend (3) in the previous subsection, we derived the colors 
of the two satellites. The colors obtained from simultaneous multicolor observations
can be defined as follows: $g'-R_{\rm C}$, $g'-I_{\rm C}$, $R_{\rm C}-I_{\rm C}$ 
(MITSuME), $r-i$, $r-z$, $i-z$ (MuSaSHI), $V-H$, $B-H$ (HONIR), $J-H$, $J-K_s$, 
and $H-K_s$ (NIC). Table 4 summarizes the averaged colors of Visorsat and 
STARLINK-1113, where, the word, ``averaged'' means the difference in average 
magnitudes (table 3) between two bands. From the system difference of optical 
and near infrared imaging, the HONIR data yields a 15 s time lag between 
two bands; the central or end exposure time, the coordinates, phase angles, angular 
velocity, and so forth are different between the $V$ (or $B$) and $H$ band (table 2). 
In this study, the colors, $V-H$ and $B-H$, are therefore pseudo colors. Moreover 
most of the colors are positive, and the $i-z$ and $J-H$ (or $H-K_s$) combinations present 
small (or negative) values compared with the other positive colors. This result 
indicates that the peak of the reflection flux lies in the wavelength range 
from the $i$ to $J$ band. No notable color differences between Visorsat 
and STARLINK-1113 and no clear phase angle dependence of the colors were 
observed (figures 4c and 4d). This result is naturally understood because the 
surface materials of the Visorsat and STARLINK-1113 would be the same.

\begin{table*}[!H]
\begin{flushleft}
\tbl{Magnitudes of Visorsat, STARLINK-1113, and reference stars.}{%
\begin{tabular}{ccccccccc}
\hline\noalign{\vskip3pt}
Date  & Reference Star & Vega Magnitude & Vega Magnitude & Normalized Magnitude$^*$ \\   
  & & (Reference Star) & (Satellite) & (Satellite)  \\
\hline  
\multicolumn{5}{c}{Target: Visorsat, Telescope/Instrument: 105 cm Murikabushi telescope/MITSuME$^\dagger$} \\ \hline 
2021 Feb 14 & HD 163080 &  $g'$: 10.96 & 6.51 $\pm$ 0.05 & 6.00 $\pm$ 0.06 \\
&  &  $R_{\rm C}$: 9.58 & 6.15 $\pm$ 0.04 & 5.64 $\pm$ 0.05 \\
&  &  $I_{\rm C}$: 8.91 & 5.70 $\pm$ 0.05 & 5.20 $\pm$ 0.05 \\
& TYC 421-1373-1 & $g'$: 10.91 & 6.41 $\pm$ 0.05 & 5.91 $\pm$ 0.06 \\
&  &  $R_{\rm C}$: 10.06 & 6.10 $\pm$ 0.04 & 5.59 $\pm$ 0.05 \\
&  &  $I_{\rm C}$: 9.62 & 5.73 $\pm$ 0.05 & 5.22 $\pm$ 0.06 \\
& & Average  ($g'$): & 6.46 $\pm$ 0.05 & 5.96 $\pm$ 0.06 \\
& & Average ($R_{\rm C}$): & 6.12 $\pm$ 0.04 & 5.62 $\pm$ 0.05 \\
& & Average ($I_{\rm C}$): & 5.72 $\pm$ 0.05 & 5.21 $\pm$ 0.06 \\
\hline 
2021 Feb 15 & TYC 1475-739-1 & $g'$: 10.74 & 7.61 $\pm$ 0.06 & 7.51 $\pm$ 0.06 \\
&  &  $R_{\rm C}$: 9.93 & 6.92 $\pm$ 0.03 & 6.82 $\pm$ 0.03 \\
&  &  $I_{\rm C}$: 9.43 & 6.15 $\pm$ 0.07 & 6.05 $\pm$ 0.07 \\
& UCAC4 558-053934 &  $g'$: 13.46 & 7.43 $\pm$ 0.17 & 7.33 $\pm$ 0.17 \\
&  &  $R_{\rm C}$: 12.11 & 6.83 $\pm$ 0.08 & 6.74 $\pm$ 0.08 \\
&  &  $I_{\rm C}$: 11.49 & 6.20 $\pm$ 0.12 & 6.10 $\pm$ 0.12 \\
& & Average  ($g'$): & 7.52 $\pm$ 0.13 & 7.42 $\pm$ 0.13 \\
& & Average ($R_{\rm C}$): & 6.88 $\pm$ 0.06 & 6.78 $\pm$ 0.06 \\
& & Average ($I_{\rm C}$): & 6.18 $\pm$ 0.10 & 6.08 $\pm$ 0.10 \\
\hline 
2021 Mar 14 & UCAC4 623-007922 &  $g'$: 11.90 & 7.77 $\pm$ 0.15 & 6.09 $\pm$ 0.15 \\
&  & $R_{\rm C}$: 10.93 & 7.16 $\pm$ 0.04 & 5.48 $\pm$ 0.04 \\
&  &  $I_{\rm C}$: 10.40 & 6.69 $\pm$ 0.06 & 5.01 $\pm$ 0.07 \\
& TYC 2332-522-1 &  $g'$: 11.36 & 7.72 $\pm$ 0.09 & 6.05 $\pm$ 0.09 \\
&  &  $R_{\rm C}$: 10.77 & 6.85 $\pm$ 0.03 & 5.17 $\pm$ 0.04 \\
&  &  $I_{\rm C}$: 10.38 & 6.53 $\pm$ 0.06 & 4.86 $\pm$ 0.07 \\
& & Average  ($g'$): & 7.74 $\pm$ 0.12 & 6.07 $\pm$ 0.12 \\
& & Average ($R_{\rm C}$): & 7.00 $\pm$ 0.03 & 5.33 $\pm$ 0.04 \\
& & Average ($I_{\rm C}$): & 6.61 $\pm$ 0.06 & 4.94 $\pm$ 0.07 \\
\hline 
\multicolumn{5}{c}{Target: STARLINK-1113, Telescope/Instrument: 105 cm Murikabushi telescope/MITSuME$^\dagger$} \\ \hline 
2021 Feb 5 & UCAC4 370-003115 & $g'$: 12.76 & 5.77 $\pm$ 0.11 & 5.00 $\pm$ 0.11 \\
& & $R_{\rm C}$: 12.01 & 5.23  $\pm$ 0.04 & 4.47 $\pm$ 0.05 \\
& & $I_{\rm C}$: 11.60 & 4.82 $\pm$ 0.09 & 4.06 $\pm$ 0.10 \\
& UCAC4 370-003114 & $g'$: 13.91 & 6.00 $\pm$ 0.20 & 5.24 $\pm$ 0.21 \\
& & $R_{\rm C}$: 13.07 & 5.09 $\pm$ 0.09 & 4.33 $\pm$ 0.10 \\
& & $I_{\rm C}$: 12.59 & 4.78 $\pm$ 0.13 & 4.02 $\pm$ 0.13 \\
& & Average  ($g'$): & 5.89 $\pm$ 0.17 & 5.12 $\pm$ 0.17 \\
& & Average ($R_{\rm C}$): & 5.16 $\pm$ 0.07 & 4.40 $\pm$ 0.08 \\
& & Average ($I_{\rm C}$): & 4.80 $\pm$ 0.11 & 4.04 $\pm$ 0.12 \\
\hline
\multicolumn{5}{c}{Target: Visorsat, Telescope/Instrument: Akeno 50 cm/MITSuME$^\dagger$} \\ 
\hline
2021 Nov 11 & BD-20 6624 & $g'$: 11.22 & 6.89 $\pm$ 0.09 & 5.61 $\pm$ 0.09 \\
 &  & $R_{\rm C}$: 10.69 & 5.79 $\pm$ 0.05 & 4.51 $\pm$ 0.05 \\
 &  & $I_{\rm C}$: 10.37 & 5.21 $\pm$ 0.12 & 3.93 $\pm$ 0.12 \\ 
 & BD-20 6621 & $g'$: 11.14 & 6.83 $\pm$ 0.11 & 5.55 $\pm$ 0.11\\
 &  & $R_{\rm C}$: 10.67 & 5.71 $\pm$ 0.06 & 4.43 $\pm$ 0.06\\ 
 &  & $I_{\rm C}$:  10.33 & 5.05 $\pm$ 0.14 & 3.77 $\pm$ 0.14 \\
 &  & Average  ($g'$) & 6.86 $\pm$ 0.10 & 5.58 $\pm$ 0.10\\
 &  & Average  ($R_{\rm C}$) & 5.75 $\pm$ 0.06 & 4.46 $\pm$ 0.06\\
 &  & Average  ($I_{\rm C}$) & 5.13 $\pm$ 0.13 & 3.85 $\pm$ 0.12 \\
\hline\noalign{\vskip3pt} 
\end{tabular}}\label{table:extramath}
\begin{tabnote}
\hangindent6pt\noindent
$^*$ Magnitudes at a height of 550 km. \\
$^\dagger$ Magnitudes estimated by equation (1).\\
$^\ddagger$ Magnitudes estimated by equation (4).\\
\end{tabnote}
\end{flushleft}
\end{table*}

\begin{table*}[!H]
\setcounter{table}{2}
\begin{flushleft}
\tbl{(Continued)}{%
\begin{tabular}{ccccccccc}
\hline\noalign{\vskip3pt}
Date  & Reference Star & Vega Magnitude & Vega Magnitude & Normalized Magnitude$^*$ \\   
  & & (Reference Star) & (Satellite) & (Satellite)  \\
\hline
\multicolumn{5}{c}{Target: Visorsat, Telescope/Instrument: Akeno 50 cm/MITSuME$^\dagger$} \\ 
\hline
2022 Jan 17 & UCAC4 399-000139 & $g'$: 12.73 &  7.69 $\pm$ 0.34  & 6.61 $\pm$ 0.34 \\
 &  & $R_{\rm C}$: 12.16 & 7.35 $\pm$ 0.15 & 6.27 $\pm$ 0.15 \\
 &  & $I_{\rm C}$: 11.80 & 6.96 $\pm$ 0.28 & 5.88 $\pm$ 0.28 \\ 
 & UCAC4 399-000135 & $g'$: 12.38 & 7.70 $\pm$ 0.19 & 6.61 $\pm$ 0.19 \\
 &  & $R_{\rm C}$: 11.89 & 7.34 $\pm$ 0.13 & 6.25 $\pm$ 0.14 \\ 
 &  & $I_{\rm C}$: 11.72 & 6.94 $\pm$ 0.21 & 5.86 $\pm$ 0.21 \\
 &  & Average  ($g'$) & 7.69 $\pm$ 0.28 & 6.61 $\pm$ 0.28 \\
 &  & Average  ($R_{\rm C}$) & 7.35 $\pm$ 0.14 & 6.26 $\pm$ 0.15 \\
 &  & Average  ($I_{\rm C}$) & 6.95 $\pm$ 0.24 & 5.87 $\pm$ 0.25 \\
\hline
\multicolumn{5}{c}{Target: STARLINK-1113, Telescope/Instrument: Akeno 50 cm/MITSuME$^\dagger$} \\ 
\hline
2021 Mar 17 & BD-02 4622 & $g'$: 11.66 & 6.93 $\pm$ 0.08 & 6.28 $\pm$ 0.08 \\
 &  & $R_{\rm C}$: 10.65 & 6.02 $\pm$ 0.04 & 5.37 $\pm$ 0.04 \\
 &  & $I_{\rm C}$: 9.97 & 5.41 $\pm$ 0.09 & 4.77 $\pm$ 0.09 \\ 
 & BD-02 4617 & $g'$: 11.28 & 6.86 $\pm$ 0.07 & 6.21 $\pm$ 0.08\\
 &  & $R_{\rm C}$: 10.58 & 6.01 $\pm$ 0.06 & 5.37 $\pm$ 0.06 \\ 
 &  & $I_{\rm C}$: 10.11 & 5.43 $\pm$ 0.12 & 4.78 $\pm$ 0.12 \\
 &  & Average  ($g'$) & 6.90 $\pm$ 0.08 & 6.25 $\pm$ 0.08 \\
 &  & Average  ($R_{\rm C}$) & 6.02 $\pm$ 0.05 & 5.37 $\pm$ 0.05 \\
 &  & Average  ($I_{\rm C}$) & 5.42 $\pm$ 0.11 & 4.77 $\pm$ 0.10 \\
\hline
2021 Mar 18 & IRAS 17251+1347 & $g'$: 11.52 & 6.01 $\pm$ 0.10 & 5.83 $\pm$ 0.10 \\
 &  & $R_{\rm C}$: 9.45 & 4.95 $\pm$ 0.02 & 4.78 $\pm$ 0.02 \\
 &  & $I_{\rm C}$: 7.72 & 4.45 $\pm$ 0.04 & 4.27 $\pm$ 0.04 \\ 
 & BD+13 3376 & $g'$: 10.00 & 5.87 $\pm$ 0.04 & 5.70 $\pm$ 0.04 \\
 &  & $R_{\rm C}$: 9.45 & 5.10 $\pm$ 0.03 & 4.93 $\pm$ 0.03 \\ 
 &  & $I_{\rm C}$: 8.95 & 4.59 $\pm$ 0.04 & 4.41 $\pm$ 0.04 \\
 &  & Average  ($g'$) & 5.94 $\pm$ 0.08 & 5.77 $\pm$ 0.08 \\
 &  & Average  ($R_{\rm C}$) & 5.03 $\pm$ 0.02 & 4.85 $\pm$ 0.03 \\
 &  & Average  ($I_{\rm C}$) & 4.52 $\pm$ 0.04 & 4.34 $\pm$ 0.04 \\
\hline
2021 Mar 22 & TYC 4151-615-1 & $g'$: 10.57 & 6.23 $\pm$ 0.01 & 5.06 $\pm$ 0.02 \\
 &  & $R_{\rm C}$: 9.91 & 5.15 $\pm$ 0.01 & 3.97 $\pm$ 0.01 \\
 &  & $I_{\rm C}$: 9.56 & 4.64 $\pm$ 0.04 & 3.49 $\pm$ 0.04 \\ 
 & TYC 4151-677-1 & $g'$: 11.71 & 6.21 $\pm$ 0.03 & 5.03 $\pm$ 0.03 \\
 &  & $R_{\rm C}$: 10.80 & 5.15 $\pm$ 0.02 & 3.98 $\pm$ 0.02 \\ 
 &  & $I_{\rm C}$: 10.31 & 4.61 $\pm$ 0.09 & 3.44 $\pm$ 0.09 \\
 &  & Average  ($g'$) & 6.22 $\pm$ 0.03 & 5.05 $\pm$ 0.03 \\
 &  & Average  ($R_{\rm C}$) & 5.15 $\pm$ 0.02 & 3.98 $\pm$ 0.02 \\
 &  & Average  ($I_{\rm C}$) & 4.63 $\pm$ 0.07 & 3.45 $\pm$ 0.07 \\
\hline
\multicolumn{5}{c}{Median magnitude of Visorsat and STARLINK-1113 in $g'$, $R_{\rm C}$, and $I_{\rm C}$ bands} \\ 
\hline
 & & Visorsat ($g'$) & 7.52 $\pm$ 0.10 & 6.07 $\pm$ 0.10 \\
 & & Visorsat ($R_{\rm C}$) & 6.84 $\pm$ 0.04 & 5.61 $\pm$ 0.05 \\
 & & Visorsat ($I_{\rm C}$) & 6.18 $\pm$ 0.09 & 5.21 $\pm$ 0.09 \\
 & & STARLINK-1113 ($g'$) & 6.11 $\pm$ 0.07 & 5.47 $\pm$ 0.08 \\
 & & STARLINK-1113 ($R_{\rm C}$) & 5.15 $\pm$ 0.03 & 4.63 $\pm$ 0.04 \\
 & & STARLINK-1113 ($I_{\rm C}$) & 4.71 $\pm$ 0.09 & 4.17 $\pm$ 0.09 \\
\hline\noalign{\vskip3pt} 
\end{tabular}}\label{table:extramath}
\begin{tabnote}
\hangindent6pt\noindent
\end{tabnote}
\end{flushleft}
\end{table*}

\begin{table*}[!H]
\setcounter{table}{2}
\begin{flushleft}
\tbl{(Continued)}{%
\begin{tabular}{ccccccccc}
\hline\noalign{\vskip3pt}
Date  & Reference Star & Vega Magnitude & Vega Magnitude & Normalized Magnitude$^*$ \\   
 & & (Reference Star) & (Satellite) & (Satellite)  \\
\hline
\multicolumn{5}{c}{Target: Visorsat, Telescope/Instrument: Kanata/HONIR$^\dagger$} \\ 
\hline
2021 Mar 15 & TYC 4805-96-1 & $V$: 10.64 & 7.06 $\pm$ 0.12 & 6.61 $\pm$ 0.13 \\ 
 &  & $H$: 9.86 & 4.66 $\pm$ 0.06 & 4.21 $\pm$ 0.07 \\
 & HD 295837 &  $V$: 11.15 & 7.06 $\pm$ 0.12 & 6.61 $\pm$ 0.12 \\
 &  &  $H$: 10.65 & 4.56 $\pm$ 0.10 & 4.10 $\pm$ 0.10 \\ 
 &  & Average ($V$) & 7.06 $\pm$ 0.12 & 6.61 $\pm$ 0.12 \\ 
 &  & Average ($H$) & 4.61 $\pm$ 0.08 & 4.15 $\pm$ 0.09\\
 \hline
2022 Jan 17 & TYC 5853-116-1 & $V$: 11.76 & 7.86 $\pm$ 0.16 & 6.82 $\pm$ 0.20 \\ 
 &  & $H$: 10.40 & 6.40 $\pm$ 0.10 & 5.27 $\pm$ 0.10 \\
 & UCAC4 350-001273 &  $V$: 11.01 & 7.95 $\pm$ 0.15 & 6.91 $\pm$ 0.20 \\
 & 2MASS 01080945-2008310 & $H$: 12.22 & 6.25 $\pm$ 0.41 & 5.12 $\pm$ 0.41 \\ 
 &  & Average ($V$) & 7.90 $\pm$ 0.16 & 6.86 $\pm$ 0.20 \\ 
 &  & Average ($H$) & 6.32 $\pm$ 0.30 & 5.20 $\pm$ 0.30 \\
\hline
2022 Jan 19 & UCAC4 529-148682 & $B$: 12.82 & 7.10 $\pm$ 0.31 & 6.60 $\pm$ 0.34 \\ 
 &  & $H$: 9.56 & 4.94 $\pm$ 0.08 &  4.35 $\pm$ 0.09  \\
 & UCAC4 529-148667 &  $B$: 13.91 & 7.44 $\pm$ 0.57 & 6.94 $\pm$ 0.58 \\
 &  &  $H$: 10.58 & 4.78 $\pm$ 0.10 & 4.20 $\pm$ 0.11 \\ 
 &  & Average ($B$) & 7.27 $\pm$ 0.46 & 6.77 $\pm$ 0.48 \\ 
 &  & Average ($H$) & 4.86 $\pm$ 0.09 & 4.27 $\pm$ 0.10 \\
\hline
\multicolumn{5}{c}{Target: STARLINK-1113, Telescope/Instrument: Kanata/HONIR$^\dagger$} \\ 
\hline
2021 Mar 19 & BD+29 3349 & $V$: 10.06 & 5.46 $\pm$ 0.09 & 5.16 $\pm$ 0.12 \\ 
 &  & $H$: 7.97 & 3.78 $\pm$ 0.03 & 3.48 $\pm$ 0.09 \\
 & TYC 2120-535-1 &  $V$: 11.86 & 5.64 $\pm$ 0.12 & 5.34 $\pm$ 0.15 \\
 &  &  $H$: 10.09 & 3.93 $\pm$ 0.07 & 3.63 $\pm$ 0.11 \\ 
 &  & Average ($V$) & 5.56 $\pm$ 0.11 & 5.25 $\pm$ 0.13 \\ 
 &  & Average ($H$) & 3.85 $\pm$ 0.05 & 3.55 $\pm$ 0.10 \\
\hline
2021 Mar 22 & TYC 4619-103-1 & $V$: 10.49 & 6.56 $\pm$ 0.09 & 5.32 $\pm$ 0.10\\
& TYC 4619-147-1 & $V$: 11.23 & 6.60 $\pm$ 0.09 & 5.37 $\pm$ 0.10\\
& & Average ($V$) & 6.58 $\pm$ 0.09 & 5.34 $\pm$ 0.10 \\ 
\hline
2022 Jan 24 & BD+42 4606 & $B$: 11.15 & 7.10 $\pm$ 0.13 & 6.27 $\pm$ 0.19 \\ 
 &  & $H$: 5.97 & 4.35 $\pm$ 0.04 & 3.42 $\pm$ 0.06 \\
 & TYC 3229-1917-1 &  $B$: 12.62 & 6.78 $\pm$ 0.40 & 5.95 $\pm$ 0.43 \\
 &  &  $H$: 10.75 & 4.36 $\pm$ 0.09 & 3.43 $\pm$ 0.10 \\ 
 &  & Average ($B$) & 6.94 $\pm$ 0.30 & 6.11 $\pm$ 0.33 \\ 
 &  & Average ($H$) & 4.35 $\pm$ 0.07 & 3.43 $\pm$ 0.08 \\
\hline
\multicolumn{5}{c}{Median magnitude of Visorsat and STARLINK-1113 in $V$ and $H$ bands} \\ 
\hline
 & & Visorsat ($V$) & 7.46 $\pm$ 0.14 & 6.72 $\pm$ 0.16 \\
 & & Visorsat ($H$) & 4.86 $\pm$ 0.10 & 4.28 $\pm$ 0.10 \\
 & & STARLINK-1113 ($V$) & 6.10 $\pm$ 0.09 & 5.33 $\pm$ 0.11 \\
 & & STARLINK-1113 ($H$) & 4.14 $\pm$ 0.06 & 3.45 $\pm$ 0.09 \\
\hline
\multicolumn{5}{c}{Target: Visorsat, Telescope/Instrument: Nayuta/NIC$^\ddagger$} \\ \hline 
2021 Feb 15 & UCAC4 471-058903 & $J:$ 11.31 & 4.50 $\pm$ 0.24 & 3.80 $\pm$ 0.24\\
& & $H:$ 10.71 & 4.08 $\pm$ 0.05 & 3.38 $\pm$ 0.06 \\
& & $K_s:$ 10.41 & 4.36 $\pm$ 0.10 & 3.65 $\pm$ 0.11\\
\hline\noalign{\vskip3pt} 
\end{tabular}}\label{table:extramath}
\begin{tabnote}
\hangindent6pt\noindent
\end{tabnote}
\end{flushleft}
\end{table*}

\begin{table*}[!H]
\setcounter{table}{2}
\begin{flushleft}
\tbl{(Continued)}{%
\begin{tabular}{ccccccccc}
\hline\noalign{\vskip3pt}
Date  & Reference Star & Vega Magnitude & Vega Magnitude & Normalized Magnitude$^*$ \\   
  & & (Reference Star) & (Satellite) & (Satellite)  \\
\hline
\multicolumn{5}{c}{Target: STARLINK-1113, Telescope/Instrument: Nayuta/NIC$^\ddagger$} \\
\hline
2021 Nov 2 & UCAC4 462-062055 & $J:$ 11.45 & 5.16 $\pm$ 0.08 & 3.88 $\pm$ 0.09\\
& & $H:$ 10.89 & 4.80 $\pm$ 0.04 & 3.52 $\pm$ 0.05 \\
& & $K_s:$ 10.80 & 4.84 $\pm$ 0.07 & 3.56 $\pm$ 0.07 \\
& UCAC4 462-062051 & $J:$ 10.55 & 4.80 $\pm$ 0.06 &  3.52 $\pm$ 0.07 \\
& & $H:$ 9.78 & 4.70 $\pm$ 0.03 & 3.42 $\pm$ 0.04\\
& & $K_s:$ 9.60 & 4.81 $\pm$ 0.07 & 3.53 $\pm$ 0.07\\
& & Average ($J$) & 4.98 $\pm$ 0.07 & 3.70 $\pm$ 0.08 \\
& & Average ($H$) & 4.75 $\pm$ 0.03 & 3.47 $\pm$ 0.05 \\
& & Average ($K_s$) & 4.83 $\pm$ 0.07 & 3.55 $\pm$ 0.07 \\
\hline 
\multicolumn{5}{c}{Target: Visorsat, Telescope/Instrument: 0.55 m SaCRA telescope/MuSaSHI$^\ddagger$} \\ \hline 
2021 Mar 11 & BD+61 273 & $r$: 9.60 & 8.18 $\pm$ 0.07 & 7.03 $\pm$ 0.08 \\
& & $i$: 9.39 & 7.79 $\pm$ 0.05 & 6.64 $\pm$ 0.06 \\
& & $z$: 9.28 & 7.47 $\pm$ 0.08 & 6.32 $\pm$ 0.09 \\
& TYC 4031-453-1 & $r$: 10.22 & 8.07 $\pm$ 0.07 & 6.91 $\pm$ 0.08 \\
& & $i$: 10.30 & 7.80 $\pm$ 0.05 & 6.65 $\pm$ 0.06 \\
& & $z$: 10.35 & 7.59 $\pm$ 0.08 & 6.44 $\pm$ 0.09\\
& & Average ($r$) & 8.12 $\pm$ 0.07 & 6.97 $\pm$ 0.08 \\
& & Average ($i$) & 7.80  $\pm$ 0.05 & 6.64 $\pm$ 0.06 \\
& & Average ($z$) & 7.53 $\pm$ 0.08 & 6.38 $\pm$ 0.09 \\
\hline
2021 Mar 15 & TYC 5316-793-1 &  $r$: 11.27 & 7.87 $\pm$ 0.04 & 6.56 $\pm$ 0.04 \\
& & $i$: 11.08 & 7.63 $\pm$ 0.06 & 6.32 $\pm$ 0.06\\
& & $z$: 10.98 & 7.57 $\pm$ 0.11 & 6.26 $\pm$ 0.11 \\
& TYC 5316-685-1 & $r$: 11.71 & 7.87 $\pm$ 0.04 & 6.56 $\pm$ 0.04 \\
& & $i$: 11.53 & 7.55 $\pm$ 0.06 & 6.24 $\pm$ 0.06 \\
& & $z$: 11.44 & 7.54 $\pm$ 0.11 & 6.23 $\pm$ 0.11\\
&& Average ($r$) & 7.87 $\pm$ 0.04 & 6.56 $\pm$ 0.04 \\
&& Average ($i$) & 7.59 $\pm$ 0.06 & 6.28 $\pm$ 0.06 \\
&& Average ($z$) & 7.55 $\pm$ 0.11 & 6.24 $\pm$ 0.11 \\
\hline 
2021 Nov 15  & HD 348402 & $r$: 9.58 & 5.64 $\pm$ 0.02 & 4.89 $\pm$ 0.02 \\
&&$i$: 9.27&  5.50 $\pm$ 0.02 & 4.75 $\pm$ 0.02\\
&&$z$: 9.10&  5.39 $\pm$ 0.02 & 4.64 $\pm$ 0.02 \\
& HD 348403 & $r$: 10.10 & 5.66 $\pm$ 0.02 & 4.91 $\pm$ 0.02 \\
&&$i$: 9.74&  5.52 $\pm$ 0.02 & 4.77 $\pm$ 0.02 \\
&&$z$: 9.55& 5.41 $\pm$ 0.02 & 4.66 $\pm$ 0.02 \\
&& Average ($r$) & 5.65 $\pm$ 0.02 & 4.90 $\pm$ 0.02 \\
&& Average ($i$) & 5.51 $\pm$ 0.02 & 4.76 $\pm$ 0.02 \\
&& Average ($z$) & 5.40 $\pm$ 0.02 & 4.65 $\pm$ 0.02 \\
\hline 
2021 Nov 17 & TYC 3492-1641-1 & $r$: 10.56 & 7.51 $\pm$ 0.12 & 6.31 $\pm$ 0.12 \\
&& $i$: 10.47 & 7.28 $\pm$ 0.12 & 6.08 $\pm$ 0.12 \\
&& $z$: 10.42 & 7.09 $\pm$ 0.08 &  5.89 $\pm$ 0.08 \\
& UCAC4 678-059331 & $r$: 12.74 &  7.53 $\pm$ 0.13 & 6.33 $\pm$ 0.13 \\
&& $i$: 12.61 & 7.35 $\pm$ 0.12 &  6.14 $\pm$ 0.12 \\
&& $z$: 12.54 &  7.16 $\pm$ 0.08 &  5.96 $\pm$ 0.08 \\
&& Average ($r$) & 7.52 $\pm$ 0.12 & 6.32 $\pm$ 0.12 \\
&& Average ($i$) & 7.31 $\pm$ 0.12 & 6.11 $\pm$ 0.12\\
&& Average ($z$) & 7.13 $\pm$ 0.08 & 5.92 $\pm$ 0.08 \\
\hline
\multicolumn{5}{c}{Median magnitude of Visorsat in $r$, $i$, and $z$ bands} \\ 
\hline
 & & Visorsat ($r$) & 7.70 $\pm$ 0.06 & 6.44 $\pm$ 0.06 \\
 & & Visorsat ($i$) & 7.44 $\pm$ 0.06 & 6.19 $\pm$ 0.06 \\
 & & Visorsat ($z$) & 7.31 $\pm$ 0.08 & 6.09 $\pm$ 0.09 \\
\hline\noalign{\vskip3pt} 
\end{tabular}}\label{table:extramath}
\begin{tabnote}
\hangindent6pt\noindent
\end{tabnote}
\end{flushleft}
\end{table*}

\begin{table*}[!H]
\setcounter{table}{2}
\begin{flushleft}
\tbl{(Continued)}{%
\begin{tabular}{ccccccccc}
\hline\noalign{\vskip3pt}
Date  & Reference Star & Vega Magnitude & Vega Magnitude & Normalized Magnitude$^*$ \\   
  & & (Reference Star) & (Satellite) & (Satellite)  \\
 \hline 
\multicolumn{5}{c}{Target: STARLINK-1113, Telescope/Instrument: 0.55 m SaCRA telescope/MuSaSHI$^\ddagger$} \\ 
\hline 
2022 Jan 26 & BD+15 1039 & $r$: 10.00 & 5.18 $\pm$ 0.02 & 4.25 $\pm$ 0.03 \\
 & & $i$: 9.61 & 4.71 $\pm$ 0.02 & 3.78 $\pm$ 0.03 \\ 
 & & $z$: 9.39 & 4.47 $\pm$ 0.02 & 3.54 $\pm$ 0.02 \\
 & TYC 1313-716-1 & $r$: 10.03 & 5.19 $\pm$ 0.02 & 4.26 $\pm$ 0.03 \\
&  & $i$: 9.68 & 4.73 $\pm$ 0.02 & 3.80 $\pm$ 0.03 \\
& & $z$: 9.49 & 4.52 $\pm$ 0.02 & 3.59 $\pm$ 0.02 \\
& & Average ($r$) & 5.19 $\pm$ 0.02 & 4.26 $\pm$ 0.03 \\
& & Average ($i$) & 4.72 $\pm$ 0.02 & 3.79 $\pm$ 0.03 \\
&  & Average ($z$) & 4.50 $\pm$ 0.02 & 3.57 $\pm$ 0.02 \\
\hline\noalign{\vskip3pt} 
\end{tabular}}\label{table:extramath}
\begin{tabnote}
\hangindent6pt\noindent
\end{tabnote}
\end{flushleft}
\end{table*}

\begin{table*}[!H]
\tbl{Averaged Colors of Visorsat and STARLINK-1113.}{%
\begin{tabular}{cccccccc}
\hline\noalign{\vskip3pt}
\multicolumn{5}{c}{Telescope/Instrument: 105cm Murikabushi telescope/MITSuME} \\ \hline 
Satellite  &  Date & $g'-R_{\rm C}$ & $g'-I_{\rm C}$ & $R_{\rm C}-I_{\rm C}$ \\   
\hline
Visorsat & 2021 Feb 14 & 0.34 $\pm$ 0.06  & 0.74 $\pm$ 0.07 & 0.41 $\pm$ 0.06\\   
Visorsat& 2021 Feb 15 & 0.64 $\pm$ 0.14 & 1.34 $\pm$ 0.16 & 0.70 $\pm$ 0.11 \\
Visorsat& 2021 Mar 14 & 0.74 $\pm$ 0.13 & 1.13 $\pm$ 0.14 & 0.39 $\pm$ 0.07 \\
STARLINK-1113 & 2021 Feb 5 & 0.72 $\pm$ 0.18 & 1.08 $\pm$ 0.20 & 0.36 $\pm$ 0.13 \\
\hline
\multicolumn{5}{c}{Telescope/Instrument: Akeno 50 cm/MITSuME} \\  \hline
Satellite & Date & $g'-R_{\rm C}$ & $g'-I_{\rm C}$ & $R_{\rm C}-I_{\rm C}$ \\
\hline
Visorsat & 2021 Nov 11 & 1.11 $\pm$ 0.11 & 1.73 $\pm$ 0.16 & 0.62 $\pm$ 0.14 \\  
Visorsat & 2022 Jan 17 & 0.35 $\pm$ 0.31 & 0.74 $\pm$ 0.37 & 0.39 $\pm$ 0.28 & \\ 
STARLINK-1113& 2021 Mar 17 & 0.88 $\pm$ 0.09 & 1.48 $\pm$ 0.13 & 0.60 $\pm$ 0.12 \\
STARLINK-1113 & 2021 Mar 18 & 0.91 $\pm$ 0.08 & 1.42 $\pm$ 0.09 & 0.51 $\pm$ 0.05 \\
STARLINK-1113 & 2021 Mar 22 & 1.07 $\pm$ 0.03 & 1.59 $\pm$ 0.07 & 0.53 $\pm$ 0.07 \\ \hline
\multicolumn{5}{c}{Target: Visorsat, Telescope/Instrument: Nayuta/NIC} \\ \hline 
Satellite & Date & $J-H$ & $J-K_s$ & $H-K_s$ \\ \hline
Visorsat & 2021 Feb 15 & 0.42 $\pm$ 0.24 & 0.15 $\pm$ 0.26 & $-$0.27 $\pm$ 0.11 \\
STARLINK-1113 & 2021 Nov 2 & 0.23 $\pm$ 0.08 & 0.15 $\pm$ 0.10 & $-0.08$ $\pm$ 0.07  \\
\hline 
\multicolumn{5}{c}{Target: Visorsat, Telescope/Instrument: Kanata/HONIR} \\ \hline 
Satellite & Date & $V-H$ & $B-H$ & \\ \hline
Visorsat & 2021 Mar 15 & 2.39 $\pm$ 0.11 & --- & \\ 
Visorsat & 2022 Jan 17 & 1.58 $\pm$ 0.34 & --- &  \\
Visorsat& 2022 Jan 19 & --- & 2.41 $\pm$ 0.47 &  \\
STARLINK-1113 & 2021 Mar 19 & 1.64 $\pm$ 0.14 & --- &  \\
STARLINK-1113 & 2022 Jan 24 & ---  & 2.58 $\pm$ 0.31 & \\
\hline 
\multicolumn{5}{c}{Telescope/Instrument: 0.55 m SaCRA telescope/MuSaSHI} \\ \hline 
Satellite & Date & $r-i$ & $r-z$ & $i-z$ \\
\hline 
Visorsat & 2021 Mar 11 & 0.33 $\pm$ 0.09 & 0.59 $\pm$ 0.11 & 0.27 $\pm$ 0.10 \\
Visorsat & 2021 Mar 15 & 0.28 $\pm$ 0.07 & 0.32 $\pm$ 0.12 & 0.04 $\pm$ 0.13 \\
Visorsat & 2021 Nov 15 & 0.14 $\pm$ 0.02 & 0.25 $\pm$ 0.02 & 0.11 $\pm$ 0.02 \\
Visorsat & 2021 Nov 17 & 0.21  $\pm$ 0.17  & 0.40  $\pm$ 0.15 & 0.19 $\pm$ 0.15 \\
STARLINK-1113 & 2022 Jan 26 & 0.47 $\pm$ 0.03 & 0.69 $\pm$ 0.02 & 0.22 $\pm$ 0.03 \\
\hline\noalign{\vskip3pt} 
\end{tabular}}\label{table:extramath}
\begin{tabnote}
\hangindent6pt\noindent
\end{tabnote}
\end{table*}

\begin{figure*}
 \begin{flushleft}
   \includegraphics[height=8cm,width=16cm]{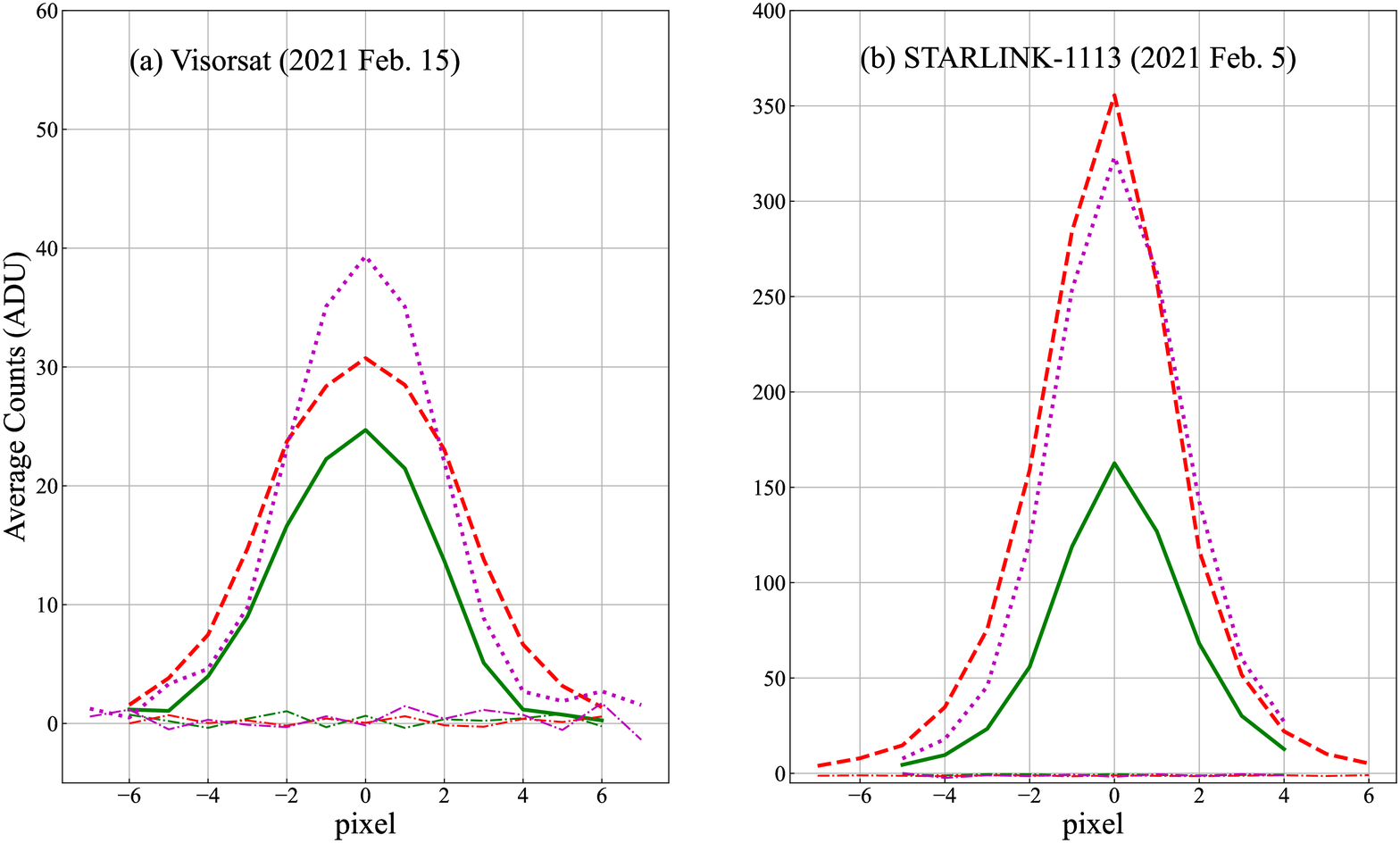} 
 \end{flushleft}
\caption{Examples of pixel-wide average section counts of (a) Visorsat and (b) 
STARLINK-1113 trails captured by the Murikabushi telescope/MITSuME $g'$ 
(solid lines), $R_{\rm C}$ (dashed lines), and $I_{\rm C}$ (dotted lines) passbands. 
Dashed-dotted lines present ADU counts of the sky region in the vicinity of the trails. }
\end{figure*}

\begin{figure*}
\begin{flushleft}
\includegraphics[height=12cm,width=16cm]{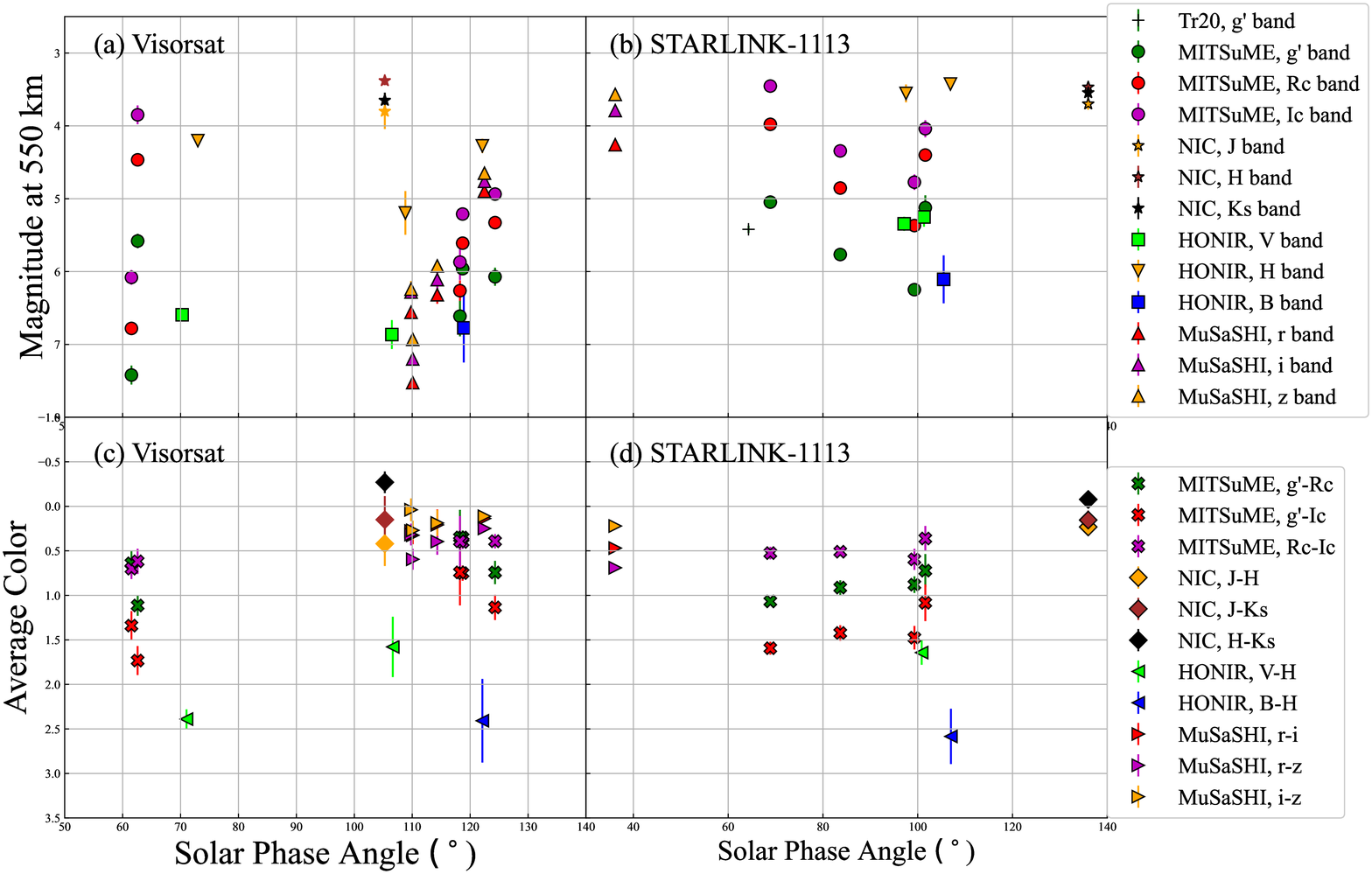} 
\end{flushleft}
\caption{Phase angle dependence of normalized magnitude and colors 
(panels a and c: Visorsat, panels b and d: STARLINK-1113). The cross in 
panel (b) indicates the $g'$-band magnitude from \citet{2020A&A...637L...1T}.} 
\end{figure*}

\section{ Discussion }
This section discusses the following contents: the reflectivity of Visorsat and 
STARLINK-1113 based on a blackbody model and a brightness comparison with 
Visorsat and Darksat.

\subsection{ The reflectivity of Visorsat and STARLINK-1113 }

\subsubsection{ Construction of the reflected flux model }
\citet{2020ApJ...905....3H} modeled the $g'$, $R_{\rm C}$, 
and $I_{\rm C}$ band flux of Darksat and STARLINK-1113 streaks, together 
with their blackbody radiation:
\begin{eqnarray}
B(\lambda,~T) = \frac{2hc^2}{\lambda^5}\frac{1}{\exp{(\frac{hc}{kT\lambda})}-1},
\end{eqnarray}
where $h$, $\lambda$, $c$, and $k$ are the Planck constant, wavelength, 
the speed of light, and the Boltzmann constant, respectively.
This study evaluated the Vega magnitudes of Visorsat and STARLINK-1113. 
In this section, we convert these magnitudes to AB flux with the zero-point 
correction by \citet{2007AJ...133..734B} and then fit it to the blackbody model. 

We define the covering factor, $C_{\rm f}$, as the fraction of the sun visor of a 
Visorsat that reduces the amount of sunlight reflected from the satellite surface. The 
covering factor of the Visorsat depends on  the positional relation between the observer, 
sun, and the satellite. To derive the albedo and the covering factor of the Visorsat, 
we formulated the reflection of sunlight, $F_{\rm RS}$, the earthshine, $F_{\rm REs}$, 
the thermal radiation of the satellite, $F_{\rm TS}$, and the reflection of Earth’s thermal 
radiation, $F_{\rm TE}$ (see also 
\cite{2003A&A...397..325A}) in the blackbody model: 
$\footnotesize$
\begin{eqnarray}
F_{\rm RS} &=& \pi\biggr(\frac{R_{\odot}}{1~{\rm au}}\biggr)^2B(\lambda,~T_{\odot})a_{\rm mod}
~p(\theta)U_{\rm f}\biggr(\frac{r_{\rm sat}}{h_{\rm T}}\biggr)^2~\frac{\lambda^2}{c}\\ 
F_{\rm REs} &=& a_{\rm E}\biggr(\frac{R_{\oplus}}{R_{\oplus}+h_{\rm T}}\biggr)^2
\biggr[1-\biggr(\frac{R_{\oplus}}{R_{\oplus}+h_{\rm T}}\biggr)^2\biggr]~
\frac{p(\phi)}{p(\theta)U_{\rm f}}~F_{\rm RS}\\
F_{\rm TS} &=& \pi\epsilon\biggr(\frac{r_{\rm sat}}{h_{\rm T}}\biggr)^2B(\lambda,~T_{\rm sat})~\frac{\lambda^2}{c} \\
F_{\rm TE} &=& \pi\epsilon\biggr(\frac{R_{\oplus}}{R_{\oplus}+h_{\rm T}}\biggr)^2B(\lambda,~T_{\rm E})
~a_{\rm mod}\biggr(\frac{r_{\rm sat}}{h_{\rm T}}\biggr)^2\frac{\lambda^2}{c},
\end{eqnarray}
$\normalsize$
where $U_{\rm f}~=~1~-~C_{\rm f}$, $T_{\odot}~(=5772~{\rm K})$ is the temperature 
of the Sun, $T_{\rm E}~(=290~{\rm K})$ is the temperature of Earth's surface,
$T_{\rm sat}~(=280~{\rm K}, $ cf. \cite{2020ApJ...905....3H}) is the temperature of the 
surface of the Starlink satellites, $R_{\odot}~(=7.0\times10^5~{\rm km})$ 
(and $R_{\oplus}~=~6371 ~{\rm km}$) is the radius of the Sun (and the Earth),  
$h_{\rm T}~(=~550~{\rm km})$ is the orbital height of the satellites, 
$a_{\rm mod}$ (and $a_{\rm E}~=~0.3$) is the modeled albedo with the blackbody
(and albedo of the Earth), $\epsilon$ is infrared emissivity ($=~0.9$; \cite{1986Icar...68..239L}), 
and $\theta$ and $\phi$ are the solar (Sun-target-observer), and observer (Sun-observer-target) 
phase angle, respectively. The phase integral, $p(\theta)$, as a first-approximation 
function is expressed by: 
\begin{eqnarray}
p(\theta) = \frac{2}{3}\biggr[\biggr(1-\frac{\theta}{\pi}\biggr)\cos\theta + \frac{1}{\pi}\sin\theta \biggr].
\end{eqnarray}
We adapted a satellite radius, $r_{\rm sat}$, of 1.69 m, based on the surface area 
of the Starlink satellites of 3 $\times$ 3 m$^2$. The derivation of equations (11) -- (14) 
is explained in the Appendix of \citet{2020ApJ...905....3H}. To restrict the aforementioned  
blackbody model, we set the following rules: (i) Visorsats and normal Starlink satellites 
have the same albedo, $a_{\rm mod}$, because the base of Visorsat uses the same surface 
material as normal Starlink satellites; (ii) the covering factor, $C_{\rm f}$, is zero for the 
ordinary Starlink satellites (i.e., STARLINK-1113 in this study); (iii) if the albedo of STARLINK-1113 
differs among several observations, the average of them is adopted as the albedo of 
Visorsat; and (iv) for simplicity, this analysis does not consider the attitude of these satellites. 

The albedo error, $\sigma_{\rm a}$, of STARLINK-1113 was calculated by applying the law 
of error propagation to the albedo as a function of $F_{\rm tot},~\theta,~\phi, ~\lambda_{\rm c}$, 
where $F_{\rm tot}$ is the sum of the left-hand sides of equations (11) -- (14) and the satellite 
flux obtained from our observations:

\begin{eqnarray}
\sigma_a = \sqrt{\biggr(\frac{\partial a}{\partial F_{\rm tot}}\delta F_{\rm tot}\biggr)^2+\biggr(\frac{\partial a}{\partial \theta}\delta \theta \biggr)^2+\biggr(\frac{\partial a}{\partial \phi}\delta \phi\biggr)^2}.
\end{eqnarray}
We computed the AB flux, $F_{\rm tot}$, and its error, $\delta F_{\rm tot}$, in Jy. 
In the cases of observations with the Murikabushi telescope/MITSuME $g'$, 
$R_{\rm C}$, and $I_{\rm C}$ bands, for example, we defined the final albedo error, 
$\sigma_{\rm a, final}$, as follows:
\begin{eqnarray}
\sigma_{\rm a, final}^2 = \frac{\sigma_{\rm a}(g')^2 + \sigma_{\rm a}(R_{\rm C})^2 + \sigma_{\rm a}(I_{\rm C})^2}{3}. 
\end{eqnarray}
In this study the errors for each band were comparable. The covering factor errors 
of Visorsat are similarly determined by the law of error propagation for a function, 
$C_{\rm f} (\bar{a},~F_{\rm tot},~\theta,~\phi;~\lambda_{\rm c})$, where $\bar{a}$ is the
averaged albedo for each epoch of STARLINK-1113.

\subsubsection{ Model fitting to AB flux }
Figures 5 -- 8 present the fitting results of the blackbody model to the AB flux at 
the height of 550 km for Visorsat and STARLINK-1113, and their albedo and covering 
factor are also shown. While the range of the covering factor of $0.18~\leq~C_{\rm f}~\leq~0.92$ 
is so wide that its average value over figures 5 -- 8 ($\bar{C_{\rm f}}~=~0.57$) is 
consistent with that of the design of Visorsat \citep{2021arXiv210706026C}. We confirmed 
that the systematic darkness of Visorsat compared with that of STARLINK-1113 (e.g., figure 4) 
can be interpreted by the covering factor (i.e., the effect of the visor on Visorsat). 

Exceptionally on 2021 March 14  (or November 15), the Visorsat trail captured with the 
Murikabushi telescope (or SaCRA telescope) showed an outstandingly high flux, to 
which we can not fit by the blackbody model with 0.10 (or 0.06) for the albedo of 
STARLINK-1113 (see figures 5d and figure 6d). The product of albedo and covering factor, 
$a_{\rm mod}U_{\rm f}$, of 0.13 (or 0.18) for Visorsat taken with the Murikabushi 
telescope (or SaCRA telescope) is the largest among the values in figure 5 (or in figure 6).
We considered the effect of moonlight reflection from Visorsat: the elevation of the nearly 
full moon on 2021 November 15 was $\sim$ 37$^\circ$ at the observation site. However, 
the model flux of the moonlight reflection (see Appendix C in \cite{2020ApJ...905....3H}) 
was found to be negligible 10$^{-5}$ Jy at most. We probably observed Visorsat located 
in the azimuth and elevation regions, where the reflection 
became highly effective \citep{2022AJ....163...21L}. The diffuse reflection from the front 
surface of the solar panel probably contributed to the high flux of the Visorsat trail captured 
by the SaCRA telescope, owing to the brightness from the back surface of the panel 
being approximately 6.0 mag even in the lower orbit, according to observations by \citet{2021arXiv210706026C}.
Other possible scenarios of unexpected outliers include differences in the satellite 
attitude; the attitude difference between satellites produces extremely bright or dark trails. 
As shown in figures 4a and 4b, there are cases where the scattering of magnitudes is 
noticeable even between datapoints with close phase angles: the covering-factor 
difference of Visorsat between the phase angles at 61.4$^\circ$ and 62.6$^\circ$ 
(figure 4a) is $\sim$ 0.6 (figures 5c and 5h). This scattering and the extremely large 
reflection flux aforementioned may be explained by the difference in satellite attitude. 
However, incorporating the satellite attitude into the blackbody model is difficult  
because TLEs do not contain such information. Hence, a quantitative discussion 
on this topic will be presented in our future work.

In both satellites, the wavelength at which the reflected components 
(i.e., $F_{\rm RS}+F_{\rm REs}$) peak is approximately in the $z$ band range, which is  
consistent with the results of the color estimation in subsection 3.1 (see table 4 and figure 6). 
The estimated albedo, $a_{\rm mod}$, in the $J,~H,$ and $K_s$ bands (= 0.47 for the Nayuta/NIC 
data, figure 8) tended to be larger than that in the optical range. This trend also appeared in the 
Kanata/HONIR data: the $H$ (or $V$ and $B$) band flux was well fitted by the blackbody models 
with albedo larger (or smaller) than the $V$ and $B$ (or $H$) band flux (figure 7). Although it is 
difficult to explain the higher albedo ($a_{\rm mod}~=~0.47$) than that in the $H$ 
band obtained with the Kanata/HONIR data ($a_{\rm mod}=0.14~-~0.20$) with the simple 
blackbody model, the albedo can be estimated to be much smaller by considering the 
reflection from the solar panels for both satellites \citep{2021arXiv210706026C}. Hence the 
longer the wavelength separation among different bands, the more difficult it is to fit the 
blackbody model with a unique albedo for the satellite flux. For example, the absolute 
reflectivities of SOLBLACK, PHOSBLACKIII\footnote{$\langle$https://www.akg.jp/puresijyon$\_$e/products/solbk.htm$\rangle$}, 
and aluminum rapidly increased above a wavelength of $\sim$ 1 $\mu$m. 
This result likely indicates the wavelength dependence of the reflectivity on 
satellite-surface materials. 

\subsection{ Relations among magnitude, phase angle and covering factor }
Another study \citep{2021arXiv210100374M} claimed  
that the visual magnitudes of Visorsat had a minimum value around a phase 
angle of 90$^\circ$. However, to discuss the detailed effect of the sun 
visor of Visorsat, we had to investigate the relations between normalized magnitude 
($m$), phase angle ($\theta$), and covering factor ($C_{\rm f}=1-U_{\rm f}$). As 
shown in figure 9, we confirmed these relations by comparing the difference between 
Visorsat and STARLINK-1113, using $\theta$ - $aU_{\rm f}$ and $aU_{\rm f}$ - $m$ 
distributions. We estimated the Spearman's rank correlation coefficient for the 
$\theta$ - $aU_{\rm f}$ distribution of Visorsat as follows:
\begin{eqnarray}
r_{\rm s} = 1-\frac{6\sum^{N}_{i=1}d_i^2}{N^3-N},
\end{eqnarray}
where $N$ and $d_i$ are the number of data points of the two variables and the
difference in the ranks of the $i$-th element, respectively. The magnitude scattering 
of the MITSuME data is likely caused by the phase angle or albedo dependence 
of their magnitudes (figure 9d). Moreover, the albedo of STARLINK-1113 
is almost independent of the phase angle in the same wavelength range (figure 9c). 
Therefore, the upward-sloping distributions for Visorsat (figures 9a and 9b) indicate 
that (i) the covering factor is moderately anti-correlated with respect to the phase angle 
($r_{\rm s} = ~0.495$ with $p$ = 0.072); and (ii) the smaller the covering factor is, the brighter 
the magnitudes of Visorsat tend to become. The same trend was observed when we 
considered the satellite radius, $r_{\rm sat}$ as a parameter ($r_{\rm sat}\geq1.69\times\alpha$ m 
with $\alpha\geq1.0$, i.e., including the solar panels) based on figure 21 in 
\citet{2021arXiv210706026C}. Thus, we quantitatively demonstrated the the shading effect 
of the visor on Visorsat and also verified the phase angle dependence of this effect under 
conditions far from Earth's shadow. In figure 9a, two outliers, on the left side (one 
of the MITSuME data) and the NIC data, suggest that factors other than the phase 
angle reduce how effective the shading effect of the sun visor is (e.g., orientation of 
the visor along our line of sight). Nevertheless, the normalized magnitudes tended to 
decrease as the covering factor increased (figure 9b). Because the $J$, $H$, and $K_s$ 
band magnitudes of Visorsat were relatively bright (table 3), those data points were 
plotted in the upper right of figure 9b. 

\subsection{ Comparison of Visorsat and Darksat magnitudes }
Although we demonstrated the shading effect of Visorsat, its impact on astronomical 
observations remains serious. Regarding the normalized $g'$-band magnitude of 
Visorsat, the median value of 6.07 $\pm$ 0.10 mag (table 3) is comparable to or 
slightly dimmer than that of 5.7 mag estimated by \citet{2022AJ....163..199B}. 
In addition, these magnitudes are slightly brighter than those of Darksat in the $g'$ band
 (e.g., \cite{2020A&A...637L...1T}, $\sim$ 6.1 mag; \cite{2020ApJ...905....3H}, 
$\sim$ 6.3 mag). The measurement by \citet{2021A&A...647A..54T} demonstrated that 
the normalized magnitudes of Darksat in the $r'$, $i'$, $J$, and $K_s$ bands were 
5.63 $\pm$ 0.07 mag, 5.00 $\pm$ 0.03 mag, 4.21 $\pm$ 0.01 mag, and 3.97 $\pm$ 
0.02 mag, respectively. In table 3, the median magnitudes of Visorsat are significantly 
dimmer (or slightly brighter) than those of Darksat in the $r$ and $i$ bands (or in the 
$J$ and $K_s$ bands): 6.44 $\pm$ 0.06 mag (for the $r$ band), 6.19 $\pm$ 0.06 mag 
(for $i$ band), 3.80 $\pm$ 0.24 mag (for $J$ band), and 3.65 $\pm$ 0.11 mag (for the
$K_s$ band). On the basis of the aforementioned, concluding that 
Visorsat is darker than Darksat is difficult. Visorsat-designed satellites should to 
implement new countermeasures to reduce reflected sunlight.

\begin{figure*}
 \begin{flushleft}
   \includegraphics[height=21cm,width=16cm]{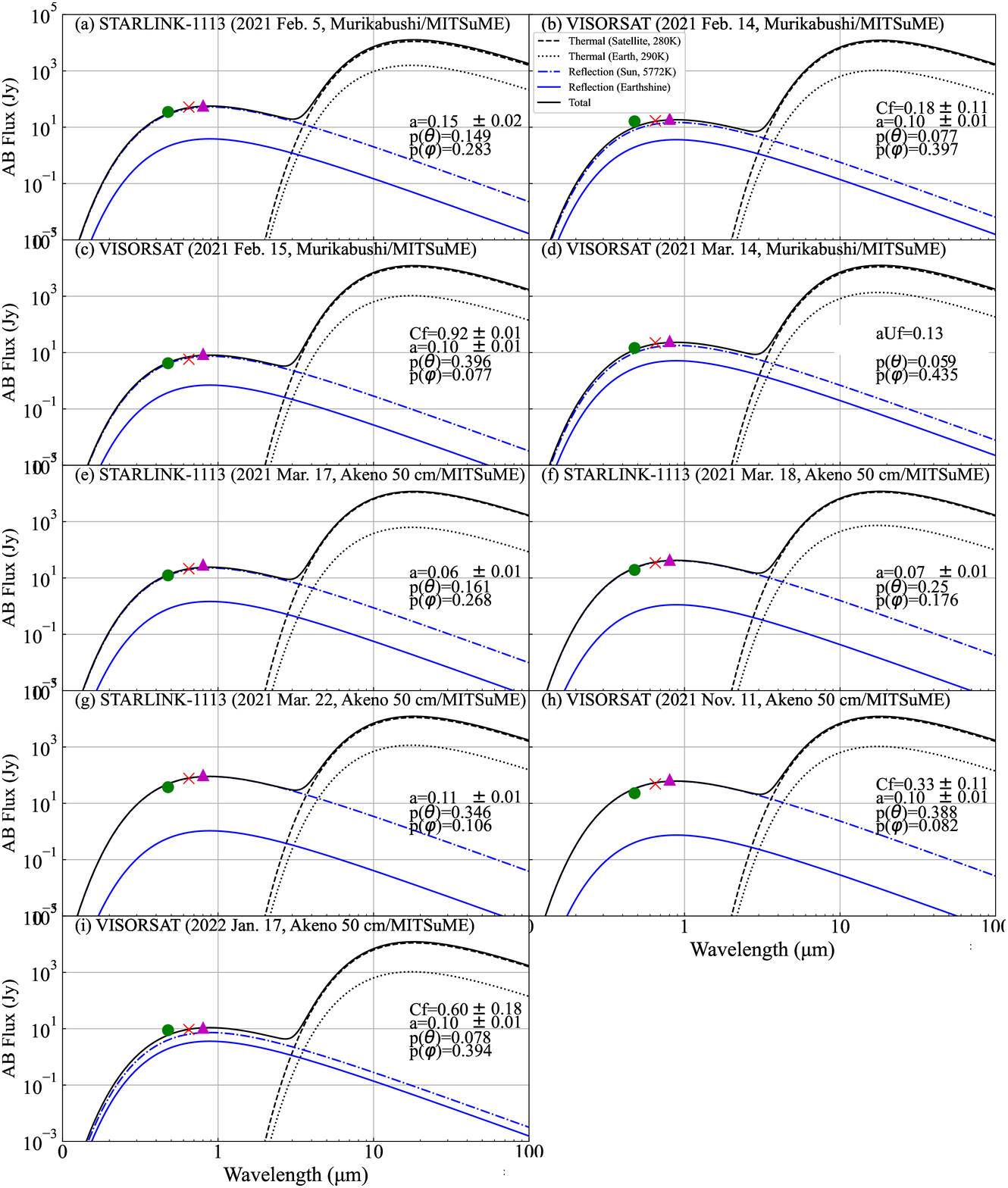} 
 \end{flushleft}
\caption{Blackbody radiation model for the $g'$- (circles), $R_{\rm C}$- (crosses), 
and $I_{\rm C}$- (triangles) bands AB flux of Visorsat and STARLINK-1113 captured with the 
Murikabushi and Akeno 50 cm telescope/MITSuME. Black solid lines show the 
total flux of the reflection of the sunlight (blue dashed-dotted lines), the earthshine 
(blue chain double-dashed lines), the thermal radiation from the satellites (black 
dashed lines), and the reflection of Earth's thermal radiation (black dotted lines). 
The phase angle, $p(\theta)$, and $p(\phi)$, and albedo, $a_{\rm mod}$, 
(or the covering factor of the sun visor of Visorsat, $C_{\rm f}$) are 
described in each panel (or panels b, c, d, h, and i). The albedo of 0.10 for Visorsat is 
the average value of observations for STARLINK-1113 (i.e.,  panels a, e, f, and g ).}
\end{figure*}

\begin{figure*}
 \begin{flushleft}
 \includegraphics[height=15cm,width=16cm]{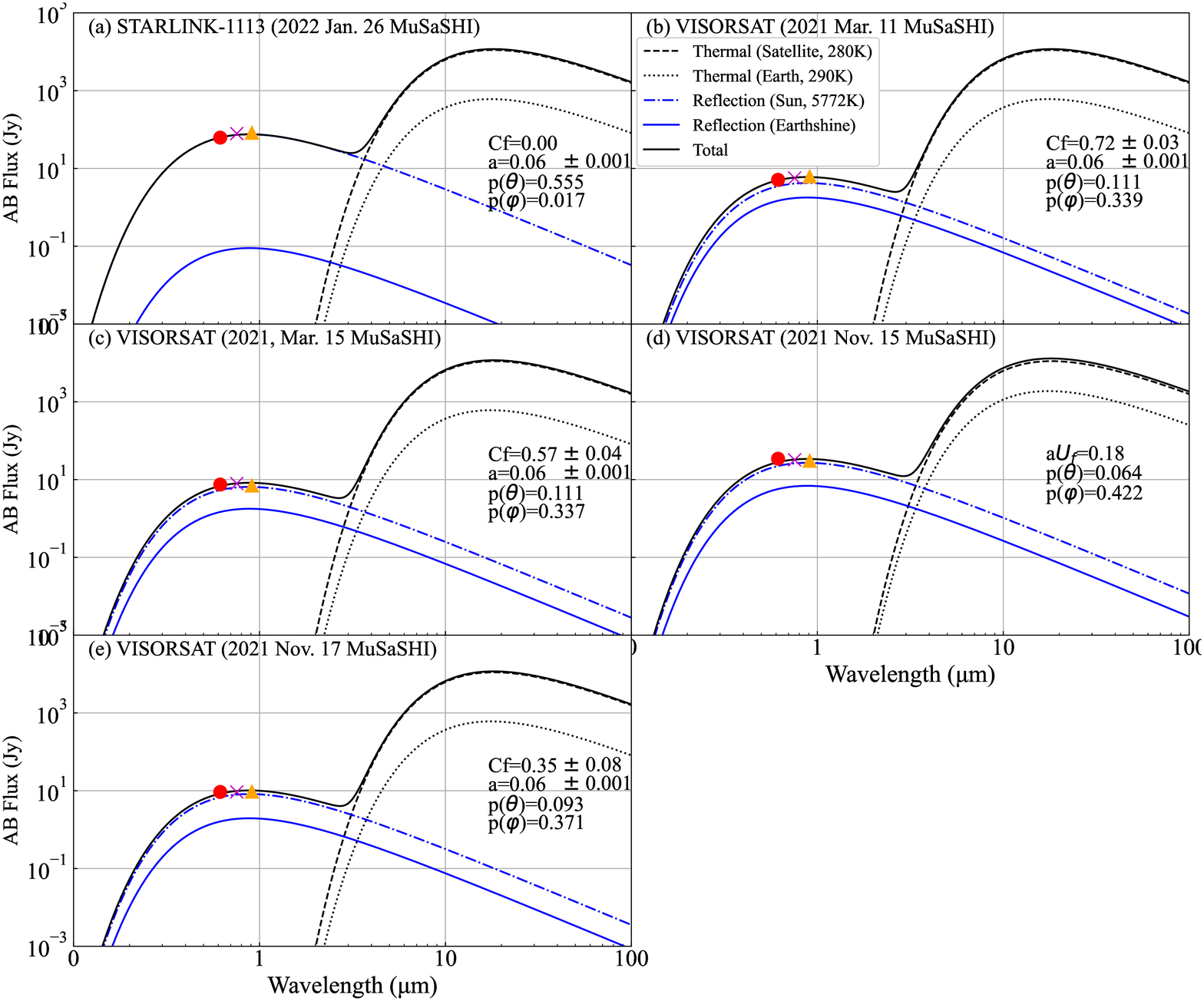} 
 \end{flushleft}
\caption{Same as figure 5 but for $r$ (circles), $i$ (crosses), and $z$ (triangles) 
bands captured with the SaCRA telescope/MuSaSHI.}
\end{figure*}

\begin{figure*}
 \begin{center}
   \includegraphics[height=21cm,width=15cm]{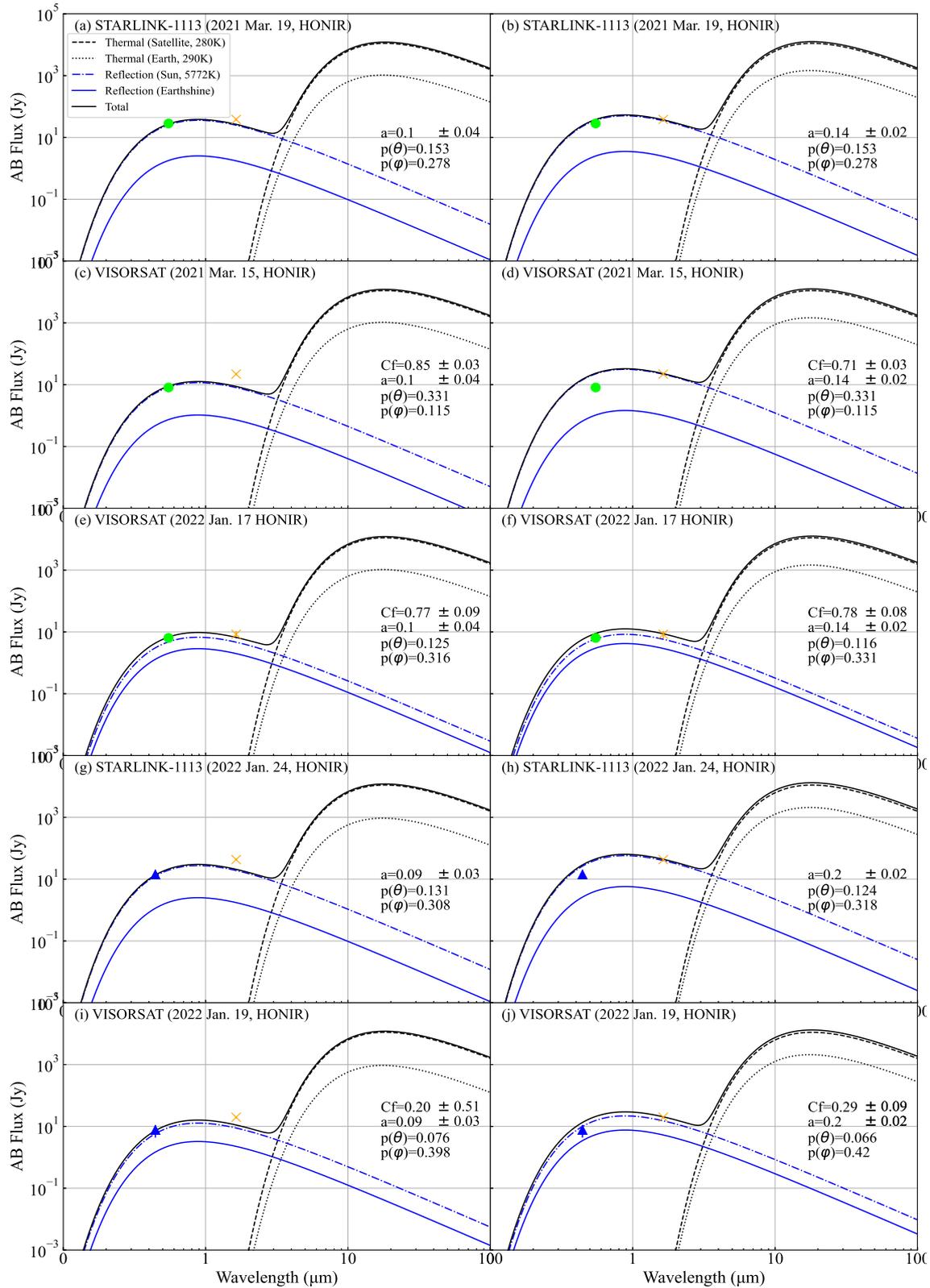} 
 \end{center}
\caption{Same as figure 5 but for $V$ (circles), $B$ (triangles), and 
$H$ (crosses) bands captured with the Kanata telescope/HONIR. Because 
the albedo considerably differs between the $V$ (or $B$) and $H$ 
bands, the model fitting to AB flux is performed separately for 
each band.}
\end{figure*}

\begin{figure*}
\begin{flushleft}
\includegraphics[height=7cm,width=16cm]{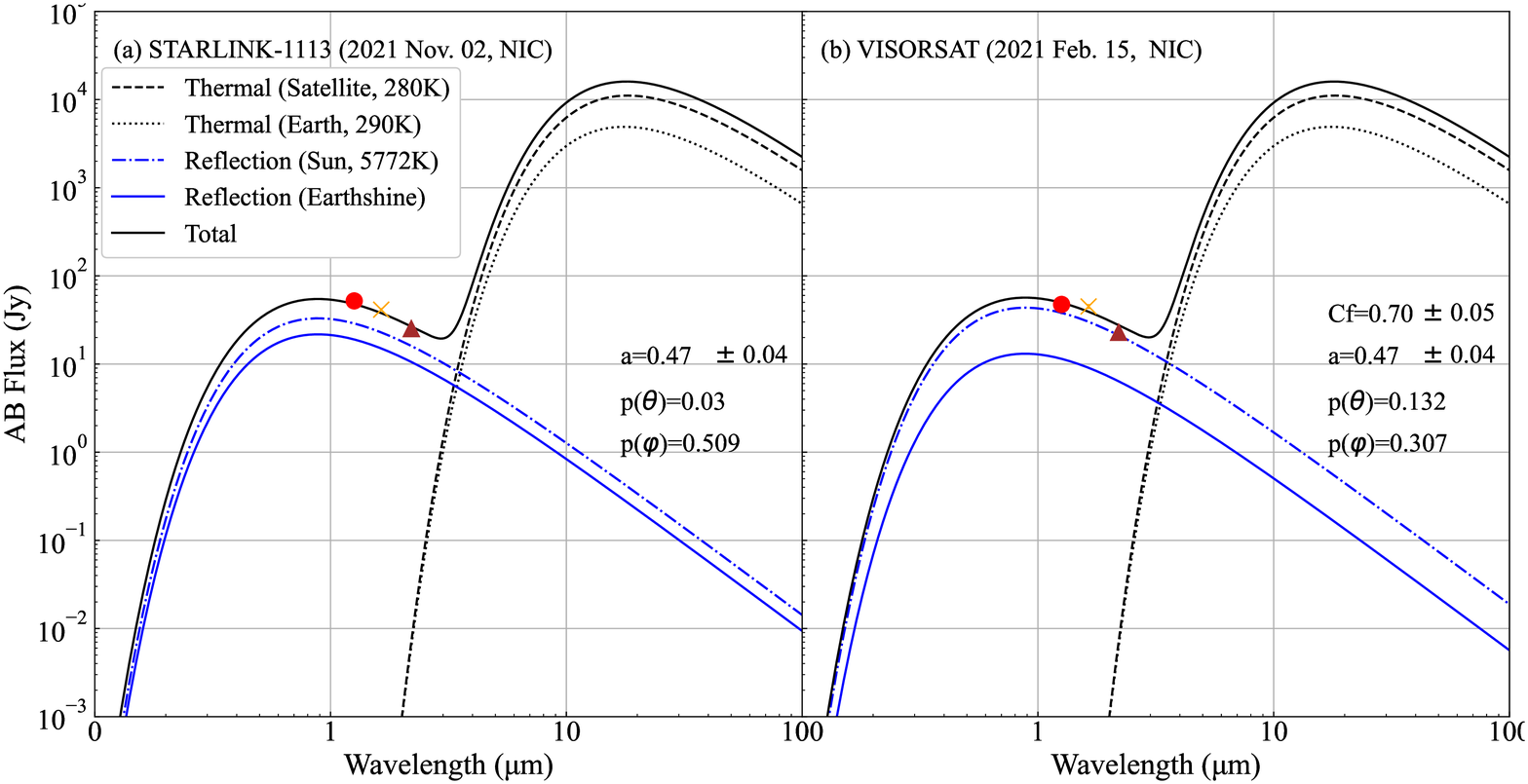} 
 \end{flushleft}
\caption{Same as figure 5 but for $J$ (circles), $H$ (crosses), and $K_s$ (triangles) 
bands taken captured the Nayuta telescope/NIC.}
\end{figure*}

\begin{figure*}
 \begin{flushleft}
   \includegraphics[height=10.5cm,width=16cm]{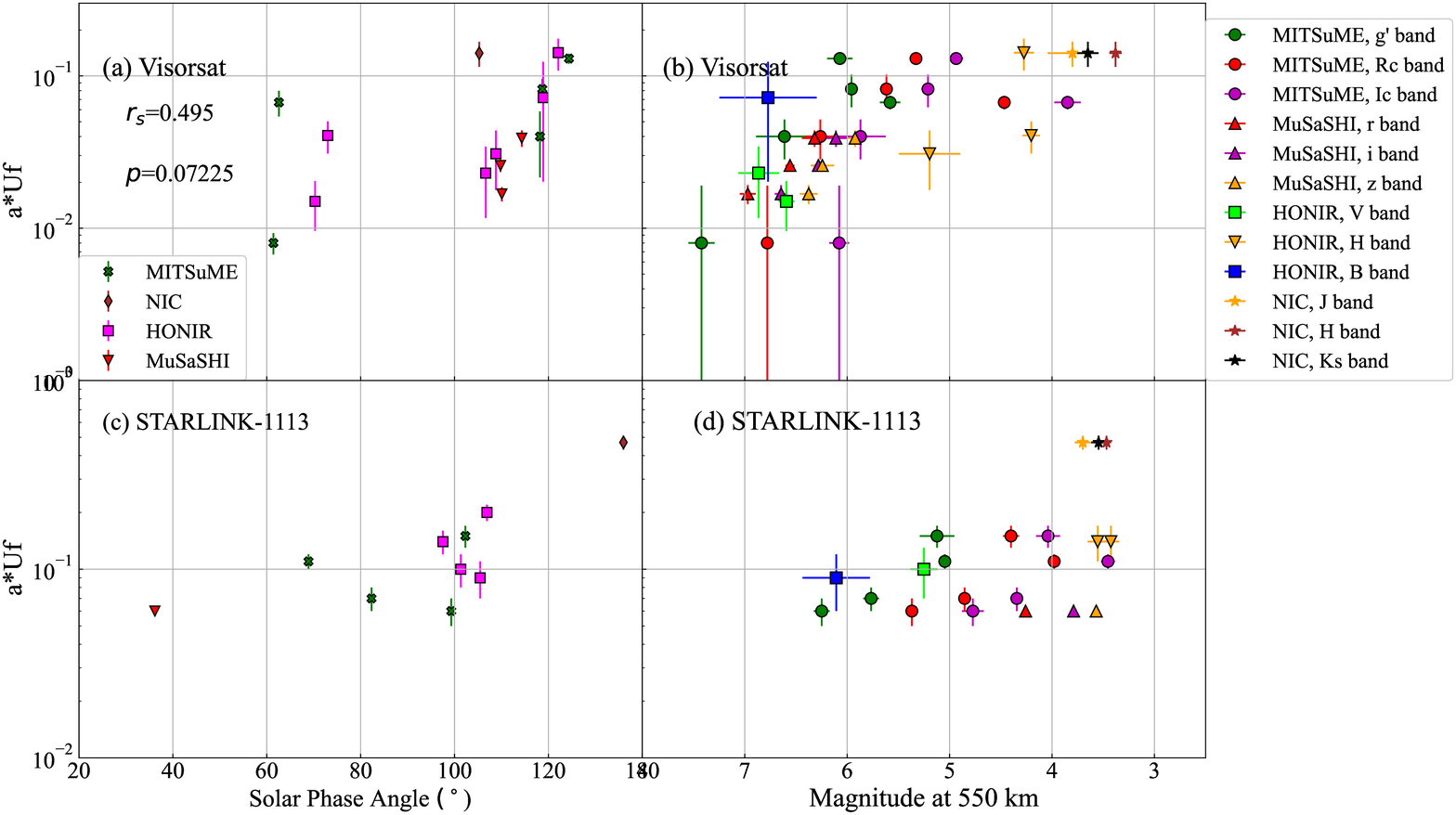} 
\end{flushleft}
\caption{Phase angle dependence of $a_{\rm mod}U_{\rm f}$ of (a) Visorsat and 
(c) STARLINK-1113, and $a_{\rm mod}U_{\rm f}$ dependence of the normalized 
magnitudes of (b) Visorsat and (d) STARLINK-1113. In panels (c) and (d), $U_{\rm f}$ is 
fixed at 1.0. Spearman's rank correlation coefficient, $r_{\rm s}$ and its $p$ value are 
described in panel (a).}
\end{figure*}

\section{ Conclusion }
We measured the multicolor magnitudes of Virorsat and STARLINK-1113 with 
simultaneous multicolor observations by the OISTER collaboration and investigated 
the effects of the sun visor of Virorsat. The results of this study can be summarized 
as follows: 

\begin{itemize}
  \item[(1)] In most cases, Virorsat is dimmer than STARLINK-1113, and the 
   sunshade on Visorsat therefore contribute to the reduction of reflected sunlight 
   (subsections 3.1, 4.1, and 4.2; $C_{\rm f}~>~0$);
  \item[(2)] The normalized magnitudes of both satellites often reach the 
   naked-eye limiting magnitude ($<$ 6.0); 
  \item[(3)] From the color estimation (subsection 3.1 and table 4) and the blackbody 
  radiation model analysis (subsection 4.1), the peak of the reflected components 
  of both satellites are around the $z$ band range (figure 6);
   \item[(4)] The albedo in the near infrared range is larger than 
   that in the optical (figures 7 and 8);
    \item[(5)] For considering the shading effect of Visorsat, the multivariable analysis 
    between the phase angle, the magnitude at 550 km altitude, and the covering factor 
    show that two parameters are necessary: the phase angle and the orientation of the 
    visor along our line of sight (figure 9).
\end{itemize}
Thus, the simultaneous multicolor observations facilitated our detailed investigation 
of a shading treatment for Visorsat. However, the observational impact from 
Visorsat remains profound. Visorsat-designed satellites are required to further reduce the 
reflected sunlight with new countermeasures. For example, if Visorsat is equipped with 
visors on both sides, the sunlight reflection will be reduced further. 

\section*{ Acknowledgment }
We thank to the staff at the Ishigakijima Astronomical Observatory and Yaeyama Star 
Club for the supporting the authors (T.H. and H.H). M.O. is also grateful to all the staff 
at the Public Relations Center, NAOJ. This research is supported by the 
Optical and Infrared Synergetic Telescopes for Education and Research (OISTER) 
program funded by the MEXT of Japan. We are indebted to Takahiro Nagayama, 
Kouji Ohta, Kazuhiro Sekiguchi and Mamoru Doi for their helpful comments. 
The anonymous referee provided very useful comments for improving this study. 
The fourth U.S. Naval Observatory CCD Astrograph Catalog (UCAC4) was provided 
by Zacharias N., Finch C.T., Girard T.M., Henden A., Bartlet J.L., Monet D.G., and 
Zacharias M.I. This publication uses data products from the Two Micron All Sky 
Survey, which is a joint project of the University of Massachusetts and the Infrared 
Processing and Analysis Center/California Institute of Technology, funded by the 
National Aeronautics and Space Administration and the National Science Foundation. 
The MITSuME system was supported by a Grant-in-Aid for Scientific Research on 
Priority Areas (grant number 19047003).

\title{}
\author{list of authors} 
\altaffiltext{}{the authors' affiliation}
\KeyWords{}

\maketitle

\end{document}